\documentclass[12pt]{iopart}
\usepackage{amsfonts}
\usepackage{amssymb}
\usepackage{latexsym}
\usepackage{graphicx}
\usepackage{dcolumn}
\usepackage{bm}
\usepackage{textcomp}
\usepackage{wasysym}
\usepackage{stmaryrd}
\usepackage{subfig}

\def \ee{\end{equation}}
\def \be{\begin{equation}}
\def \bea{\begin{eqnarray}}
\def \eea{\end{eqnarray}}

\newcommand{\eqref}[1]{(\ref{#1})}

\newcommand{\cO}{\mathcal{O}}
\newcommand{\cR}{\mathcal{R}}

\newcommand{\m}{\mu}
\newcommand{\n}{\nu}
\newcommand{\p}{\partial}

\newcommand{\gb}{\bar{g}}

\newcommand{\I}{i}

\begin{document}

\title{
Structural aspects of asymptotically safe black holes
}

\author{Benjamin Koch* and Frank Saueressig**}
 \address{
*Pontificia Universidad Cat\'{o}lica de Chile, \\
Av. Vicu\~{n}a Mackenna 4860, \\
Santiago, Chile \\

**Radboud University Nijmegen, \\ Institute for Mathematics, Astrophysics and Particle Physics (IMAPP),\\
Heyendaalseweg 135, 6525 AJ Nijmegen, The Netherlands \\
}
\date{\today}

\begin{abstract}
We study the quantum modifications of classical, spherically symmetric 
Schwarzschild (Anti-) de Sitter black holes within Quantum Einstein Gravity. The quantum effects are incorporated through the running coupling constants $G_k$ and $\Lambda_k$, computed within the exact renormalization group approach, and a common scale-setting procedure. We find that, in contrast to common intuition, it is actually the cosmological constant that determines the short-distance structure of the RG-improved black hole: in the asymptotic UV the structure of the quantum solutions is universal and given by the classical Schwarzschild-de Sitter solution, entailing a self-similarity between the classical and quantum regime. As a consequence asymptotically safe black holes evaporate completely and no Planck-size remnants are formed. Moreover, the thermodynamic entropy of the critical Nariai-black hole is shown to agree with the microstate count based on the effective average action, suggesting that the entropy originates from quantum fluctuations around the mean-field geometry.
\end{abstract}

\maketitle



%
\section{Introduction}

Black holes \cite{Carroll:2004st,Wald:1995yp} are perhaps the most fascinating objects known to exist in the universe. In particular, they provide an important testing ground for physics involving strong gravitational fields. Within classical general relativity, the simplest class of black holes are given by a Schwarzschild black hole, embedded in either Minkowski- or (Anti-)de Sitter space. These arise as static, spherical symmetric solutions of Einsteinís equations in the presence of a cosmological constant and are completely characterized by the value of Newtonís constant, the cosmological constant and their mass, appearing as a free constant of integration. As a common feature the solutions share a space-like singularity at the center of the black hole.

Understanding the dynamics of the black hole spacetime when quantum effects of the geometry are switched on, constitutes an important challenge in theoretical physics. In particular, they provide an important laboratory for testing gravitational theories that go beyond classical general relativity. At the semi-classical level this has led to the discovery of exciting phenomena including the evaporation of a black hole due to Hawking radiation and, more general, the development black hole thermodynamics. It also nurtured the idea that a full quantum treatment of a black hole could resolve the curvature singularity lurking at its center.

Clearly, questions concerning the resolution of the black hole singularity, the final state of the black hole evaporation or the microscopic origin of the black hole entropy should be addressed within a consistent theory of quantum gravity. In this work we will investigate these questions within the effective average action approach \cite{mr} to Weinberg's Asymptotic Safety scenario \cite{Weinberg:1979}, see \cite{Niedermaier:2006wt,Reuter:2007rv,Percacci:2007sz,Litim:2008tt,Reuter:2012id} for current reviews. The key ingredient in this scenario is a non-Gaussian fixed point (NGFP) of the gravitational RG-flow which provides the UV-completion of gravity at trans-Planckian energies. By now, the properties of this fixed point have been studied in a series of
works first approximating $\Gamma_k$ by the Einstein-Hilbert action \cite{Dou:1997fg,Souma:1999at,Lauscher:2001ya,Reuter:2001ag,Litim:2003vp} and subsequently adding higher derivative terms \cite{Lauscher:2001rz,Lauscher:2002sq,Rechenberger:2012pm,Codello:2007bd,Machado:2007ea,Bonanno:2010bt,Falls:2013bv,Christiansen:2012rx} and the square of the Weyl-tensor \cite{Benedetti:2009rx,Benedetti:2009gn,Niedermaier:2009zz}. Furthermore, quantum effects in the ghost sector have been studied in \cite{Eichhorn:2009ah} and RG-flows including surface terms have been considered in \cite{Becker:2012js}. As a key common result, all these works found a NGFP of the gravitational theory space suitable for Asymptotic Safety. Following \cite{Reuter:2012id}, we will
call the resulting quantum theory of gravity Quantum Einstein Gravity (QEG).

 The phenomenological consequences of Asymptotic Safety are most conveniently explored based on the gravitational effective average action $\Gamma_k[g_{\mu\nu}]$ \cite{mr}. In a nutshell, $\Gamma_k[g_{\mu\nu}]$ is constructed in such a way that, when evaluated at tree level, it describes gravitational phenomena, \emph{including quantum corrections}, whose typical momenta are of the order $k$.  Subsequently, this property has been used to study quantum corrections to classical space-times by including the effect of scale-dependent coupling constants either at the level of the classical solutions, classical equations of motion, or the Lagrangian density and subsequently identifying $k$ with a physical cutoff scale of the system. Starting from the pioneering works \cite{Bonanno:1998ye,Bonanno:2000ep}, these techniques have been used to explore quantum-modifications of classical Schwarzschild black holes \cite{Emoto:2005te,Bonanno:2006eu,Ward:2006vw,Falls:2010he,Basu:2010nf,Contreras:2013hua} and black holes including angular momentum \cite{Reuter:2006rg,Reuter:2010xb}. The inclusion of higher-derivative terms has been considered in \cite{Cai:2010zh} while the reverse-engineering of the RG-improved black hole geometry based on thermodynamical properties has been undergone in \cite{Falls:2012nd}. Finally, the creation of asymptotically safe black holes in gravitational theories with large extra dimensions has been studied in \cite{Hewett:2007st,Litim:2007iu,Koch:2007yt,Burschil:2009va} and conditions on the formation of a singularity based on the RG-improved collapse of a thin matter shall have been derived \cite{Casadio:2010fw}. These works find that the NGFP underlying Asymptotic Safety indeed has significant consequences for the emergent structure of the quantum-improved black holes. For the Schwarzschild case, the quantum corrections either soften the spatial singularity encountered in the interior of the black hole or remove it entirely \cite{Bonanno:2000ep}. Even more striking, the quantum-improvement significantly affects the semi-classical black hole evaporation process. Instead of evaporating completely, the temperature of the quantum-improved black hole drops to zero at a finite mass, suggesting that the final state of the black hole evaporation is a Planck-mass remnant \cite{Bonanno:2000ep}.

The novel feature of the present work is the inclusion of a running cosmological constant in the RG-improvement procedure. This step is inevitable in a consistent implementation of the RG-flow: even if one sets $\Lambda_k = 0$ at one fixed scale $k$, the RG-flow will generate a non-vanishing cosmological constant dynamically. In particular the dimensionless cosmological constant at the NGFP is positive, $\lambda_*$, entailing that the dimensionful counterpart $\Lambda_k$ will diverge quadratically at high energies. Thus it is a priori unclear that setting $\Lambda_k = 0$, an approximation that underlies all previous RG-studies of black holes, is a good approximation. In fact, we will show that including the running of $\Lambda$ drastically modifies of the picture of black holes in Asymptotic Safety. While counter-intuitive at first sight, the cosmological constant has a crucial effect on resolution of the black hole singularity and the black hole evaporation process. Moreover, it opens up new insights on the origin of the black hole entropy within Asymptotic Safety.

The rest of this work is organized as follows. Sects.\ \ref{Sect.2} and \ref{Sec2} review the properties of classical, spherical symmetric black hole solutions incorporating a non-trivial cosmological constant and the RG-flow of QEG in the Einstein-Hilbert truncation, respectively. This discussion provides the basis for introducing our RG-improvement procedure in Sect.\ \ref{sect:2.3}. The properties of the resulting quantum-improved black hole solutions are investigated in Sects.\ \ref{SecUVasympt} (short distance structure), \ref{SecUVIR} (global structure of the solutions), and \ref{sect.7} (thermodynamic properties), respectively. In Sect.\ \ref{sec.sum} we summarize our results by highlighting the
improvement-scheme dependent and independent conclusions. Appendix A finally contains complementing information on improved
actions and improved equations of motions.

\section{Classical black hole solutions with cosmological constant}
\label{Sect.2}
We start with the discussion of classical static and spherical symmetric black hole solutions in the presence of 
a cosmological constant. These solutions are well-known and correspond to the line-element
\be\label{lineele}
ds^2= -f(r) \, dt^2+ f(r)^{-1} \, dr^2 + r^2 d\Omega_2^2
\ee
with
\be\label{frfct}
f(r) = 1 - \frac{2 G M}{r} - \frac{1}{3} \, \Lambda \, r^2 \, . 
\ee
Here $G$ and $\Lambda$ denote the classical Newton constant and cosmological constant, $d\Omega_2^2$ is
the volume element of the two-sphere, and $M$ is an integration constant. Depending on the sign of $\Lambda$,
the line-element describes a Schwarzschild-AdS ($\Lambda < 0$), Schwarzschild $(\Lambda = 0)$,
or Schwarzschild-dS ($\Lambda > 0$) black hole, respectively. 

This class of solutions possesses a space-like singularity at $r = 0$. This can be made visible
 by computing the square of the Riemann tensor,
which is independent of the choice of coordinates
\be\label{singClass}
R_{\m\n\rho\sigma} R^{\m\n\rho\sigma} = \frac{48 G^2 M^2}{r^6} + \frac{8 \Lambda^2}{3} \, .
\ee

Moreover, the black hole solutions give rise to horizons where \eqref{frfct} vanishes.
 Multiplying $f(r)$ by $r$, the roots
of the cubic equation can be found analytically,
\be\label{AdSroot}
r_{\rm 0} = - \cR^{-1/3} - \frac{1}{\Lambda} \cR^{1/3} \, , \quad
\ee
and
\be\label{dSroot}
r_{\pm} = \frac{1}{2} \left( 1 \pm \I \sqrt{3} \right) \cR^{-1/3}  + \frac{1 \mp \I \sqrt{3}}{2 \Lambda} \cR^{1/3} \, . 
\ee
Here 
\be
\cR = 3 G M \Lambda^2 + \sqrt{ 9 G^2 M^2 \Lambda^4 -\Lambda^3}\quad.
\ee
These roots constitute a black hole horizon if the resulting $r$ is real and positive. The
actual horizon structure of the black holes therefore depends on the sign of $\Lambda$
and $M$ and can be determined from the determinant of the cubic $r f(r)$. 
For $\Lambda < 0$ there is a single real root,
\be
r_{\rm AdS} \equiv r_0 \quad,
\ee
while $r_{\pm}$ constitutes a complex pair. Thus the Schwarzschild-AdS black hole
has a single black hole horizon. For $\Lambda = 0$ this horizon agrees
with the one of the Schwarzschild black hole $r_{\rm SS} = 2 G M$.

For Schwarzschild-dS black holes, $\Lambda > 0$, the situation is slightly more complicated.
For $M = 0$, the line element \eqref{lineele} reduces to dS space with
a cosmological horizon at 
\be
r_c = \sqrt{3/\Lambda} = r_+ |_{M = 0} \, . 
\ee
For $M>0$ a black hole horizon emerges at 
\be
\tilde r \equiv r_- \, . 
\ee 
For increasing $M$, the black hole horizon grows while the cosmological horizon $r_c \equiv r_+$ shrinks. 
For the critical mass
\be\label{Mcrit}
M_{\rm max} \equiv \frac{1}{3 \, G \, \sqrt{\Lambda}}
\ee
the two horizons coincide and one obtains the Nariai black hole as the maximal black hole in dS space. For $M > M_{\rm max}$ the
two horizons disappear and the line-element \eqref{lineele} describes a naked singularity.

Given a horizon situated at $r_{\rm H}$, one can determine its thermodynamic properties  
foremost the Hawking temperature
\be\label{temp}
T_H=\left.\frac{\partial_r f(r)}{4 \pi}\right|_{r=r_{\rm H}}\quad 
\ee
and the Beckenstein-Hawking entropy
\be\label{entro}
S = \frac{\pi r_{\rm H}^2}{G} \, . 
\ee
For the Schwarzschild-dS black holes, which will play a central role in our
investigation, the thermodynamics of the two horizons has been discussed extensively in the context of the dS/CFT correspondence
\cite{Cai:2001sn,Cai:2001tv}. The cosmological horizon $r_c$ has Hawking temperature and entropy
\be
T_c  =  \frac{1}{4 \pi r_c} \left( \Lambda \, r_c^2 - 1 \right) \, , \qquad 
S_c  =  \frac{\pi r_c^2}{G} \, , 
\ee
while the evaluation of \eqref{temp} and \eqref{entro} at the black hole horizon $\tilde r$ gives
\be\label{horent}
\tilde T  =  \frac{1}{4 \pi \tilde r} \left( 1 - \Lambda \, \tilde r^2 \right) \, , \qquad
\tilde S  =  \frac{\pi \tilde r^2}{G} \, . 
\ee

Remarkably, the entropy associated with both the cosmological and black hole horizon
can be expressed through a Cardy-Verlinde formula \cite{Cai:2001sn,Cai:2001tv} which
is conjectured to capture the entropy of a conformal field theory (CFT) valid
in arbitrary dimension \cite{Verlinde:2000wg}. This provides a strong hint
that the thermodynamics of the Schwarzschild-dS black hole
has the correct entropy to be described by a CFT.

For the cosmological horizon this equivalence can be shown as follows. 
Using a surface-counterterm method, the gravitational mass of the 
Schwarzschild-dS black
hole was determined in \cite{Balasubramanian:2001nb,Ghezelbash:2001vs}
\be\label{energy}
E = - M = \frac{r_c}{2 G_N} \left( \frac{1}{3} \,  \Lambda \,  r_c^2 - 1 \right) \, . 
\ee
which expresses the mass $M$ of the black hole in terms of the radius of the cosmological horizon.
The Casimir energy $E_c$, defined as the non-extensive part of \eqref{energy}, $E_c \equiv 3 E - 2 T S$ 
is given by \cite{Cai:2001sn}
\be\label{casimir}
E_c = - \frac{r_c}{G}
\ee
Given the relations \eqref{energy} and \eqref{casimir} it is straightforward to verify that the
entropy $S_c$ can be cast into Cardy-Verlinde form
\be\label{CVform1}
S_c = \pi \sqrt{\frac{3}{\Lambda}} \, \sqrt{|E_c| (2 E - E_c)} \, .
\ee

Similarly, $\tilde S$, eq.\ \eqref{horent}, also follows the Cardy-Verlinde formula, provided that one adopts the 
mass definition of Abbott and Deser \cite{Abbott:1981ff}
\be
\tilde E = M = \frac{\tilde r}{2 G} \left(1 - \frac{1}{3} \, \Lambda \, \tilde r^2 \right) \, . 
\ee
In this case the definition of the Casimir energy yields
\be
\tilde E_{c} = \frac{\tilde r}{G}
\ee
and the extensive part of the energy is
\be
2 \tilde E - \tilde E_{c} = - \frac{\Lambda}{3G} \tilde r^3 \, .
\ee
Given these quantities, it is straightforward to check that
\be\label{entbhh}
\tilde S = \pi \sqrt{\frac{3}{\Lambda}} \, \sqrt{ E_{c,+} |2 E_+ - E_{c,+}| } \,
\ee
also takes the form of a Cardy-Verlinde formula. This suggests
that both the black hole horizon and the cosmological horizon 
have the correct thermodynamics for a dual description
in terms of a dual CFT in the framework of the dS/CFT correspondence.
These results will become important when discussing the properties of the quantum-improved black holes in Sect.\ \ref{SecUVasympt}.

\section{RG-flows in Quantum Einstein Gravity}
\label{Sec2}
The Asymptotic Safety program encodes the quantum effects in a  the scale dependent gravitational effective average action $\Gamma_k[g, \xi, \bar{\xi}; \bar{g}]$,
which satisfies the exact functional renormalization group equation (FRGE) \cite{mr}
\be\label{FRGE}
\p_k \Gamma_k[g, \xi, \bar{\xi}; \bar{g}] = \frac{1}{2}  {\rm Tr}  \left[ \left( \Gamma_k^{(2)} + \cR_k \right)^{-1} \, \p_k \cR_k \right] \, .
\ee
Here $g_{\m\n}$ is the averaged metric, $\gb_{\m\n}$ an unspecified background metric, and $\xi, \bar{\xi}$ are the ghost fields.
Moreover $\Gamma_k^{(2)}$ is the second variation of $\Gamma_k$ with respect to the fluctuation fields at fixed background. The crucial ingredient
in \eqref{FRGE} is the infrared cutoff $\cR_k$ which provides a $k$-dependent mass-term for fluctuations with momenta $p^2 \gg k^2$.
It renders the trace-contribution finite and peaked at $p^2 \approx k^2$. The solutions of the FRGE interpolate between the bare action at $k \rightarrow \infty$ 
and the effective action $\Gamma[g] = \Gamma_{k=0}[g, 0, 0; \bar{g}]|_{g =\gb}$ by integrating out quantum fluctuations shell-by-shell in momentum space.
In this sense the object $\Gamma_k[g, 0, 0; \bar{g}]\equiv \Gamma_k[g]$ provides an effective description of the physics at momentum scale $k$. Its vertices 
include all quantum corrections from fluctuations with $p^2 > k^2$, while long-range fluctuations are untouched.
This property is central for using $\Gamma_k[g]$ for an effective description of the quantum physics.

While it is impossible to solve the FRGE exactly, it can still be used to gain insights on the quantum nature of physical processes
 by projecting the exact RG-flow onto subspaces spanned by a suitable ansatz for $\Gamma_k$. For the study of (A)dS black holes conducted
 in this work, we thereby restrict ourselves to the Einstein-Hilbert truncation where the gravitational part of $\Gamma_k$
 is of (euclidean) Einstein-Hilbert form 
 \be\label{action}
 \Gamma_k^{\rm grav}[g] = \frac{1}{16 \pi G_k} \int d^4x \sqrt{g} \left[-R + 2 \Lambda_k \right] 
 \ee
 and includes two scale-dependent couplings, Newton's constant $G_k$ and the cosmological constant $\Lambda_k$.
 The beta functions resulting from this truncation have first been derived in \cite{mr} and
are most conveniently expressed in terms of the dimensionless 
 coupling constants
\be\label{Gvong}
 g_k= G_k \, k^{2}\quad, \quad
\lambda_k= \Lambda_k \, k^{-2} \quad \, . 
\ee
In terms of $g_k$ and $\lambda_k$ the RG equations become a system of autonomous differential 
equations
\be\label{betaeq}
k \p_k g_k = \beta_g(g_k, \lambda_k) \, , \qquad k \p_k \lambda_k = \beta_\lambda(g_k, \lambda_k) \, ,
\ee
where 
\begin{eqnarray}\label{betafcts}
\beta_\lambda(g, \lambda) = & \, ( \eta_N - 2) \lambda 
 + \frac{1}{2\pi} g \left[ 10 \Phi^1_{2}(- 2 \lambda) - 8 \Phi^1_{2}(0) - 5 \eta_N \tilde{\Phi}^1_{2}(-2 \lambda) \right] \, , \\ 
\beta_g(g, \lambda) = & \, ( 2+\eta_N ) g \, . 
\end{eqnarray}
The anomalous dimension of Newton's constant $\eta_N$ is given by
\be\label{etan}
\eta_N(g, \lambda) = \frac{g B_1(\lambda)}{1 - g B_2(\lambda)} \, ,
\ee
with the following functions of the dimensionless cosmological constant
\begin{eqnarray}\label{Bns}
B_1(\lambda) \equiv & \,  \frac{1}{3\pi}  \Big[
5 \Phi^1_{1}(-2\lambda) - 18 \Phi^2_{2}(-2\lambda) - 4  \Phi^1_{1}(0) - 6 \Phi^2_{2}(0) \Big] \, ,\\
B_2(\lambda) \equiv & \, - \frac{1}{6\pi} \, \left[5 \tilde{\Phi}^1_{1}(-2\lambda) - 18 \tilde{\Phi}^2_{2}(-2 \lambda) \right] \, . 
\end{eqnarray}
The threshold functions $\Phi$ and $\tilde \Phi$ are defined in \cite{mr} and encode the dependence of the beta functions on the IR-regulator $\cR_k$. For 
most applications, it is convenient to work with the non-differentiable `optimized cutoff' \cite{Litim:2003vp} where
\be\label{phiopt}
\Phi^{{\rm p}}_n(w) = \frac{1}{\Gamma(n+1)} \frac{1}{1+w} \, , \qquad 
\widetilde{\Phi}^{{\rm p}}_n(w) = \frac{1}{\Gamma(n+2)} \frac{1}{1+w} \, . 
\ee

\begin{figure}[t]
\centering
\includegraphics[width=10cm]{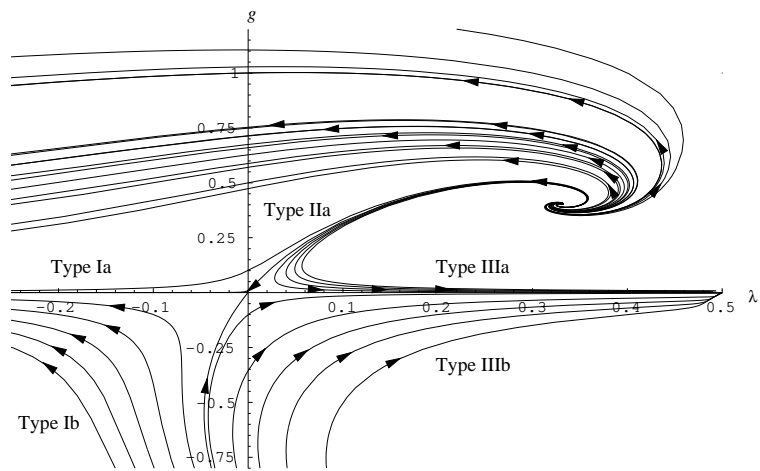}
\caption{\label{EHflow} RG flow originating from the Einstein-Hilbert truncation \eqref{betaeq} obtained with a sharp cut-off. The arrows point in the direction of increasing 
coarse-graining, i.e.\ of decreasing $k$. From \cite{Reuter:2001ag}.}
\end{figure}

The phase diagram resulting from the flow \eqref{betaeq} has been constructed in \cite{Reuter:2001ag} and is shown 
in Fig.\ \ref{EHflow}. 
The flow is governed by the interplay of a Gaussian fixed point (GFP) located at the origin, $g_* = 0, \lambda_* = 0$
and an NGFP which, for the cutoff \eqref{phiopt}, is located at 
\be\label{NGFP}
\lambda_*=0.193 \, , \qquad g_*=0.707 \, , \qquad g_* \lambda_* = 0.137\quad.
\ee
Besides the coordinates of the NGFP, we also give the value of the product $g_* \lambda_*$ which was argued to be observable \cite{Lauscher:2001ya} and its RG-scheme dependence is significantly suppressed in
comparison with the single quantities $g_*$ and $\lambda_*$ alone.

The NGFP acts as an UV-attractor for all RG-trajectories shown in the upper half-plane. Thus it governs the behavior of gravity at high energies and provides its UV-completion. 
Since the dimensionless couplings approach constant values, the NGFP also fixes the scaling behavior of the dimensionful Newton's constant and cosmological constant
at high energies. Inverting \eqref{Gvong} we obtain
\bea\label{UVG}
\lim_{k \rightarrow \infty} G_k = g_* \, k^{-2} \, , \qquad 
\lim_{k \rightarrow \infty} \Lambda_k = \lambda_* \, k^2\quad.
\eea
Following the RG-flow towards the IR there is a crossover. Depending on whether the flow passes to the left, right, or ends at the GFP, the low-energy limit is given by classical general relativity with a negative (Type Ia), positive (Type IIIa) or zero (Type IIa, Separatrix) value of the cosmological constant.

Based on the flow pattern shown in Fig.\ \ref{EHflow}, it is a plausible assumption that the NGFP will govern the structure of black holes at short distances. The universality entailed by the fixed point then lets us expect that black hole solutions which at long distances correspond to classical AdS, dS, and Schwarzschild black holes actually posses the same short distance structure. The scaling relations \eqref{Gvong} together with the fixed point condition then indicates that the value of the dimensionful cosmological constant actually becomes sizable \emph{at short distances}, indicating that in terms of RG-improved processes one should not restrict oneself to the study of Schwarzschild black holes, setting $\Lambda_k = 0$ throughout, but work in a more general class of black hole solutions which also capture the effect of the cosmological constant. 
In fact, it is this generalization that underlies the novel insights in this paper and will thoroughly modify the picture of black holes in Asymptotic Safety.

\begin{figure}[t]
   \centering
\includegraphics[width=10cm]{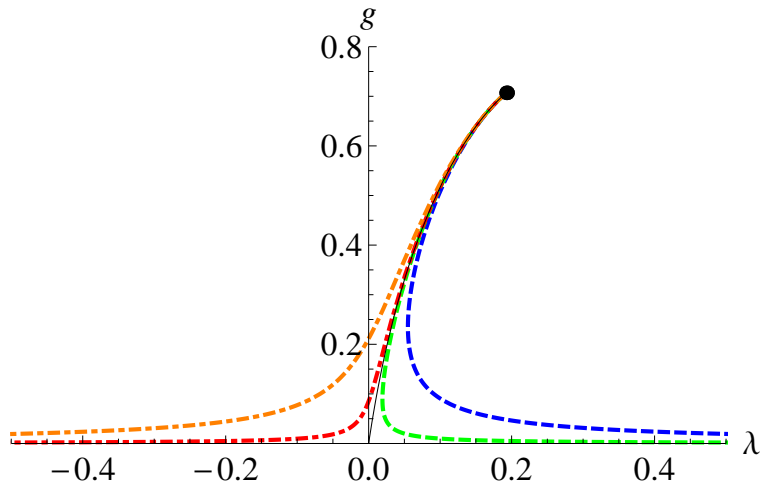}
  \caption{\label{figgVSl} 
Analytic approximation of the RG trajectories shown in Fig.\ \ref{EHflow}, provided by the parametric curves (\ref{lvong}), (\ref{gk})
for $G_0=1$, $\Lambda_0=-0.01$ (orange), $\Lambda_0=-0.001$ (red) $\Lambda_0=0$ (black), $\Lambda_0=0.001$ (green), $\Lambda_0=0.01$~(blue).
The position of the NGFP \eqref{NGFP} is marked by the black dot.}
\end{figure}

For the purpose of studying the phenomenological consequences of the RG flow depicted in Fig.\ \ref{EHflow},
it is convenient to work with approximations of the RG trajectory that can be captured by analytic expressions. Here we
will make use of the analytical curves \cite{Koch:2010nn}
\begin{eqnarray}
g(k)&=&\frac{G_0 k^2}{1+k^2/g_*} \quad,\label{gk}\\
 \lambda(g)&=&\frac{g_*\lambda_*}{g}
\left(\left(5+\frac{\Lambda_0 G_0}{g_* \lambda_*}\right)\left[1-g/g_*\right]^{3/2}-5+3g/(2g_*)(5-
g/g_*)\right) \,.\label{lvong}
\end{eqnarray}
Here the position of the NGFP \eqref{NGFP} constitutes an input parameter and the type and details of 
the RG trajectory can be dialed by specifying the IR-values of Newton's constant $G_0$ and the cosmological constant $\Lambda_0$.
For some illustrative choices of these parameters, the resulting parametric curves are shown in Fig.\ \ref{figgVSl}.
Comparing Figs.\ \ref{EHflow} and \ref{figgVSl} we readily observe that the analytically
approximated  curves describe the global 
transition from the UV to the IR. The feature that the
analytic approximation does not capture is the spiraling of the RG flow
around the NGFP.
At this stage, we stress that none of the results found in the sequel depend on the detailed properties of the RG trajectories. In particular, the modeling of the RG trajectories is not essential when studying how the gravitational 
RG flow affects classical black hole solutions at short distances.

\section{Scale setting in RG-improvement schemes}
\label{sect:2.3}
At this stage we want to exploit the information on the gravitational RG-flow
to gain insights on the structure of black holes in Asymptotic Safety. For this purpose
we apply the RG-improvement procedure, first used by Bonanno and Reuter for
classical Schwarzschild black holes \cite{Bonanno:2000ep}, to the classical
black hole solutions \eqref{lineele}. At the present stage, we limit ourselves 
to the discussion of improving the classical solutions. The case of improved
equations of motion and improved actions give rise to similar results and are 
covered in \ref{App.I}.

The basic idea underlying our RG-improvement scheme is to start from the classical
solution \eqref{lineele} and promote the coupling constants to running couplings,
$G\rightarrow G_k, \Lambda \rightarrow \Lambda_k$, whose 
scale-dependence is governed by the beta-functions \eqref{betafcts}.
The RG-improvement procedure then relates the RG-scale $k$
with a physical cutoff scale. Following the arguments in \cite{Bonanno:2000ep} this
cutoff-identification should be a coordinate invariant statement and preserve
the symmetries of the classical solution. The symmetry requirement then fixes
$k \equiv k(r)$ while the first condition of coordinate invariance
can be incorporated by constructing $k(r)$ from a geodesic quantity.
This leads to the form 
\be\label{kvonP}
k(P(r))=\frac{\xi}{d(P(r))}\quad,
\ee
where $d(P)$ is the distance scale which provides the relevant cutoff $k$ when 
the test particle is located at the point $P$. The constant
$\xi$ is expected to be of order unity and has to fixed by using 
a physics argument.
\begin{figure}[t]
  \centering
\subfloat[\hspace{7cm}.
 Functional dependence of the scale
  $k(r)/\xi$ for Schwarzschild-AdS black holes, $\Lambda_0=-0.001,\, G_0 = 1$ and $M=\{1, \,2.5, \,5, \, 10 \}$ (top to bottom curve).]{\label{figkvonrM}
\includegraphics[width=0.48\textwidth]{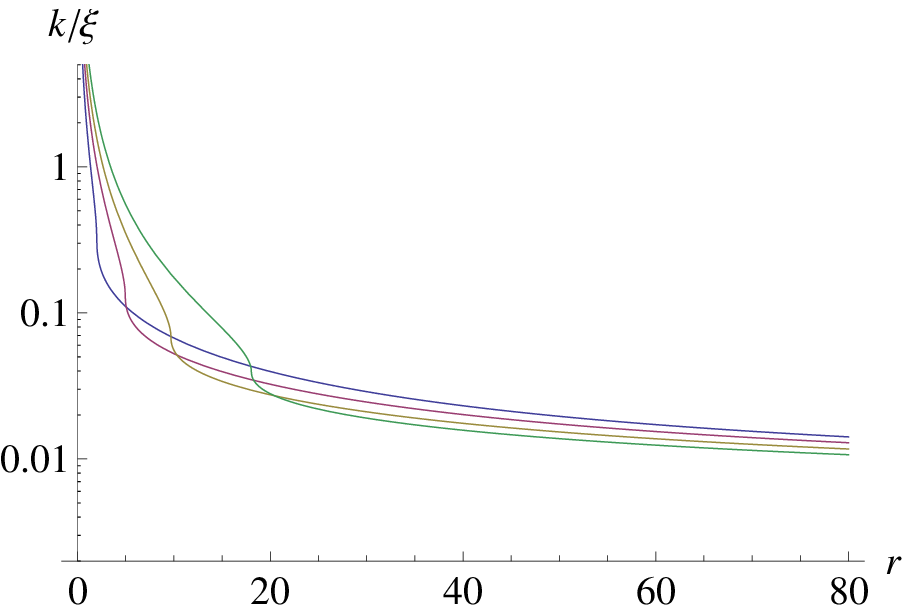}}
\hspace{0.1cm}
\subfloat[\hspace{7.0cm}.
 Functional dependence of the scale
  $k(r)/\xi$ for Schwarzschild-dS black holes with $\Lambda_0=0.001,\, G_0 = 1$ and $m=\{1,\, 2.5, \,5, \, 10 \}$ (top to bottom curve).]{\label{figkvonrP}
\includegraphics[width=0.48\textwidth]{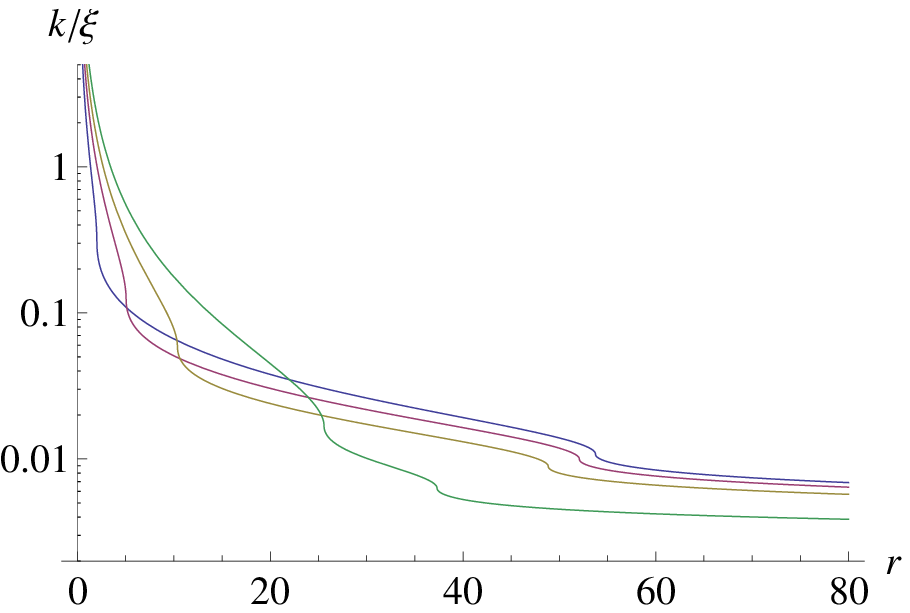}}
\caption{Cutoff-identification $k(r)$ resulting from identifying $d(r)$ with the radial geodesic distance \eqref{dproper}.}
\label{gl2}
\end{figure}
In the context of RG-improved Schwarzschild black holes, several choices for $d(r)$ have
been advocated in the literature. Here we will identify $d(P)$ with the radial proper distance
from the center of the black hole to the point $P$ along a
purely radial curve ${\mathcal{C}}_r$
\be\label{dproper}
d(r)=\int_{{\mathcal{C}}_r} \sqrt{|ds^2|} \, .
\ee
For a given classical (A)dS black hole solution, specified through $G, \Lambda, M$,
the function $d(r)$ can easily be found numerically. For two particular examples 
the resulting cutoff identification $k(r)$ is shown in Fig.\ \ref{gl2}.
The first observation is that the relation between $r$ and $k$ is
actually monotonic, i.e., large values of the momentum cutoff 
correspond to short distances and vice versa. This constitutes
an important prerequisite for the feasibility of the proposed cutoff
identification. Moreover, we observe that the cutoff-identification
contains points where the slope actually becomes infinite. These points
correspond to the horizons of the black holes, where the line-element degenerates, even though
the geodesic distance remains finite. In the AdS case the black holes have only one horizon located
at \eqref{AdSroot}. For the Schwarzschild-dS black 
holes there are two such points associated with the black hole and cosmological horizons
given by \eqref{dSroot}.  As expected from the discussion in Sect.\ \ref{Sect.2}, the two horizons
of the Schwarzschild-dS black hole approach each other for increasing mass $M$.

Besides having the full numerical cutoff-identification at our disposal,
it is also useful to obtain analytic expressions for $k(r)$ valid in the deep UV and
deep IR. For short distances, $r \rightarrow 0$
this function can be expanded in a series in $r$ whose leading terms are
easily calculated analytically
\be\label{radasym}
d(r) \simeq \frac{2}{3} \, \frac{1}{\sqrt{2 G M}} \, r^{3/2} \left( 1 + \frac{3}{10} \, \frac{r}{2 G M} +  \cO(r^{2}) \right) \, .  
\ee
The short distance behavior is completely fixed by $G$ and $M$. In particular the cosmological constant enters into the cubic term of the bracket only, and is thus 
not important in this regime. The cosmological constant is, however, crucial for the
large distance behavior of the cutoff-identification. Depending on the sign of $\Lambda$ the
 leading large-distance behavior (approximated by setting $M = 0$ in \eqref{lineele}) is
\be
d(r) \simeq 
\left\{
\begin{array}{ll}
\sqrt{\frac{3}{\Lambda}} \arcsin\left( \sqrt{\frac{\Lambda}{3}} \, r \right) \, , & \qquad \Lambda > 0 \\[1.2ex]
r \, , & \qquad \Lambda = 0 \\[1.2ex]
\sqrt{\frac{3}{-\Lambda}} {\rm arcsinh} \left(\sqrt{- \frac{\Lambda}{3}} \, r \right) \, , & \qquad \Lambda < 0\quad.
\end{array}
\right.
\ee
For $\Lambda > 0$ the geodesic distance becomes maximal at a finite value $r^2 = 3/\Lambda$ indicating the presence of the cosmological horizon. 

As an alternative to the scale-setting procedure based on the proper distance
introduced in equation (\ref{dproper}), \ref{App.0} discusses a proper-time
improvement where the cutoff is identified with the proper time measured by a freely
falling observer. This scheme has the virtue that the resulting cutoff identification
is smooth at the black hole horizon. The detailed investigation of this alternative,
carried out in \ref{App.0}, establishes that all conclusions drawn from the proper-distance improvement
remain valid when applying this alternative scale-setting procedure, pointing at the robustness of the 
results.


\section{RG-improved black hole solutions in the UV}
\label{SecUVasympt}
In this section, we will RG-improve the classical Schwarzschild-(A)dS black
hole solutions \eqref{frfct} in order to investigate how the NGFP, eq.\ \eqref{NGFP}, 
affects the short-distance structure of the black holes. Following the
discussion of Sect.\ \ref{sect:2.3}, we start from the classical solution and 
promote the coupling constants to scale-dependent quantities
\be\label{fvonrk}
f_k(r)= 1-\frac{2 M G_k}{r}-\frac{1}{3} \Lambda_k r^2 \;.
\ee
In the deep UV, $k \rightarrow \infty$, the scale dependence of the
coupling constants is governed by the NGFP. 
Substituting the fixed point scaling \eqref{UVG}, into the radial function
\eqref{fvonrk} one gets
\be\label{fvonrimp}
f_*(r) = 1-\frac{2\, M \, g_*}{k^2 \, r} - \frac{1}{3} \left( \lambda_* \, k^2 \right) r^2 \, . 
\ee 
Here $g_*$, $\lambda_*$ denote the position of the NGFP and the $*$ acts as a reminder that this solution describes the asymptotics for $k \rightarrow \infty$.

In the next step we perform the cutoff-identification \eqref{kvonP}, relating $k$ to the radial geodesic distance. 
In the UV the identification is given by
\be\label{krasym}
k(r) \simeq \frac{3}{2} \, \sqrt{2 G_0 M} \,  \xi  \, r^{-3/2} \, . 
\ee
Here $G_0$ indicates the IR-value of Newton's constant at $k = 0$. Substituting 
this asymptotics into \eqref{fvonrimp} it is straightforward to obtain the RG-improved
line-element \emph{valid at the NGFP}
\be\label{fself}
f_{*}(r) = 1-  \frac{2\, G_0 \, M}{r} \left( \frac{3}{4} \, \lambda_* \, \xi^2 \right) - \frac{1}{3} \, \left( \frac{4 \, g_*}{3 \, G_0 \, \xi^2} \right) \, r^2 \, . 
\ee   
Most remarkably, the RG-improved function $f_{*}(r)$ valid at high energies is \emph{self-similar} to the classical solution \eqref{frfct}. Promoting $k$ to a function
of $r$ interchanges the terms containing Newton's constant and the cosmological constant so that the actual $r$-dependence remains the same. Note that the inclusion
of the cosmological constant term is crucial for this self-similarity to work. 

Requiring that the classical and RG-improved solution both give rise to the
same asymptotic cutoff identification \eqref{krasym} fixes the 
numerical constant $\xi$,
\be\label{xisc}
\xi^2_{\rm sc} = \frac{4}{3 \lambda_*} \, .
\ee
For this particular value the RG-improvement scheme becomes self-consistent and we will use this criterion to fix the a priori free numerical constant $\xi$, if not 
stated otherwise. Setting $\xi = \xi_{\rm sc}$, we obtain the line-element for the self-consistent RG-improved black hole solution valid in the non Gaussian fixed point regime
\be\label{fsc}
f_{*,{\rm sc}}(r) = 1-  \frac{2 G_0 M}{r}  - \frac{1}{3} \, \Lambda_{\rm eff} \, r^2 \, , 
\ee 
with the effective cosmological constant
\be
\Lambda_{\rm eff} = \frac{g_* \, \lambda_*}{G_0} \, . 
\ee
Notably, the dimensionality of $\Lambda_{\rm eff}$ (valid in the UV) is set by the square of the Planck mass $M_{\rm Pl}^2 \equiv G_0^{-1}$, while its magnitude (in Planck units) is governed by the universal dimensionless product $g_* \, \lambda_*$. This product cannot be chosen by hand, but constitutes a prediction from Asymptotic Safety. Its magnitude has been computed in a number of works 
$g_* \, \lambda_* \approx 0.1$ \cite{Reuter:2012id,Reuter:2012xf}.

A direct consequence of the self-similarity between the classical line-element \eqref{lineele} and its RG-improved counterpart \eqref{fself} is that both metrics give rise to the same singular behavior at $r = 0$. The square of the Riemann-tensor computed from the RG-improved metric is 
\be\label{RGimprie}
R_{\m\n\rho\sigma} R^{\m\n\rho\sigma} = \frac{48 G_0^2 M^2}{r^6} \left( \frac{3}{4} \, \lambda_* \, \xi^2 \right)^2 + \frac{8}{3}  \left( \frac{4 \, g_*}{3 \, G_0 \, \xi^2} \right)^2 \, . 
\ee
Thus the RG-improvement \emph{does not resolve} the spatial singularity at $r = 0$. This is in contrast to earlier investigations of RG-improved Schwarzschild black holes carried out by Bonanno and Reuter \cite{Bonanno:2000ep,Bonanno:2006eu}, 
which reported the resolution of the singularity due to the RG improvement. 
This apparent mismatch is resolved by noting that the latter computation has essentially been carried out at $\lambda_* = 0$. Setting $\lambda_* = 0$ in \eqref{fself}, the RG-improved line-element is the one of classical de Sitter-space so that there is no singularity at the origin. This can also be confirmed from \eqref{RGimprie} which becomes regular at $r=0$ once we set $\lambda_* = 0$. At this stage, it is worthwhile stressing that the singularity appearing in this expression is not caused by the Newton's constant, but rather counter intuitively originates from the inclusion of the cosmological constant. The non-zero and positive value of the later is an intrinsic feature of the NGFP underlying Asymptotic Safety and thus cannot be avoided. 
%
\section{RG-improved black hole solutions: global structure}
\label{SecUVIR}
\begin{figure}[t]
  \centering
\subfloat[\hspace{7cm}.
 Functional dependence of the metric function
 $f(r)$ for $\Lambda_0=-0.001$, $G_0=1$, and
 from top to bottom $M=\{1,\, 2.5, \,5, \, 10\}$.
 ]{\label{figfvonrM}
\includegraphics[width=0.48\textwidth]{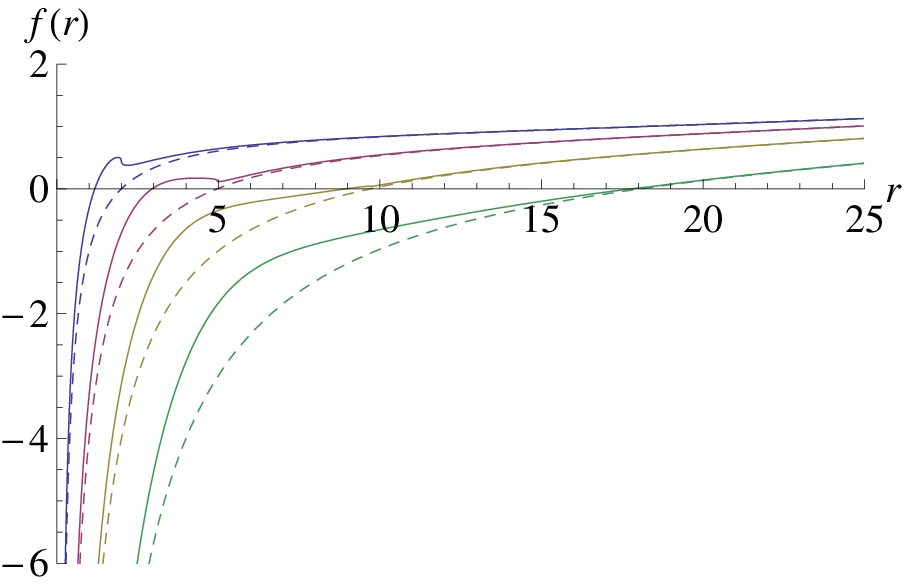}}
\hspace{0.2cm}
\subfloat[\hspace{7.0cm}.
 Functional dependence of the metric function
 $f(r)$ for $\Lambda_0=0.001$, $G_0=1$, and  from top to bottom $M=\{1,\, 2.5, \,5, \, 10\}$.
 ]{\label{figfvonrP}
\includegraphics[width=0.48\textwidth]{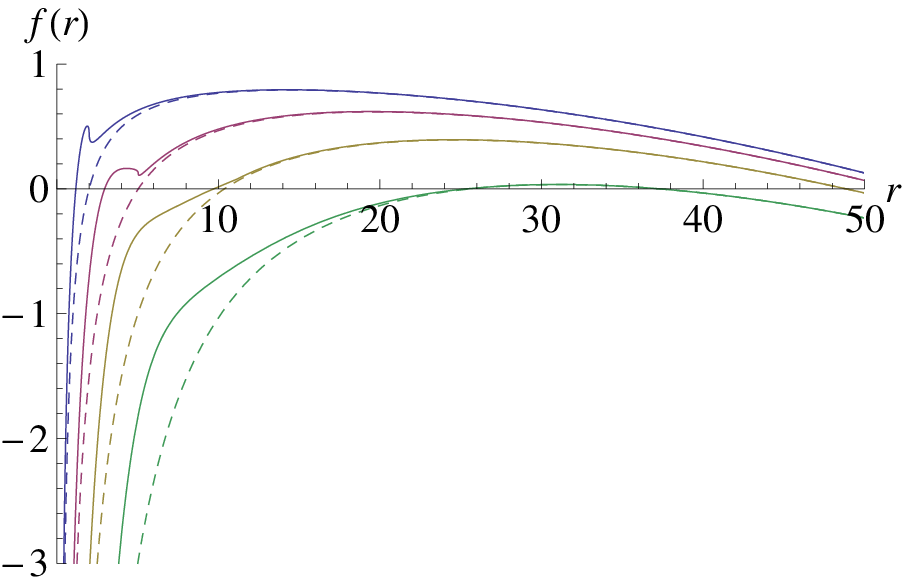}}
\caption{Radial dependence of the RG-improved metric function $f(r)$ for $\xi = \xi_{\rm sc}$   The dashed lines correspond to the classical solutions while the solid lines
 correspond to the improved solution.
 \label{figfvonr}}
\label{gl3}
\end{figure}

Following the discussion of the previous section
we will extend our investigation to the whole interval
$0 \le r < \infty$ and develop the global picture of the RG improved 
(A)dS black holes. Our starting point is again the $k$-dependent
line-element \eqref{fvonrk}. Instead of working with the
asymptotic scaling of the coupling constants where $g_k = g_*, \lambda_k = \lambda_*$, the $k$-dependence
of the dimensionless couplings is captured by the analytic approximation of 
the RG-trajectory (\ref{gk}, \ref{lvong}), however. Subsequently, 
the scale-identification is performed by replacing $k \mapsto k(r)$
with the function $k(r)$ shown in Fig.\ \ref{gl2}. The resulting
RG-improved functions $f(r)$ are then shown in Fig.\ \ref{gl3}.
Comparing the classical (dashed) and RG-improved (solid) solutions for the same values $M$, one first observes
that for $r \ll M$ and $r \gg M$ the curves actually agree with a very high precision. For small $r$ this
behavior actually reflects the self-similarity of the classical and improved solution discovered in the 
last section. The precise matching of the asymptotics thereby hinges on setting $\xi = \xi_{\rm sc}$ since
this choice also implies an agreement in the numerical coefficients fixing the short-distance behavior.  

The quantum effects captured by the RG-improvement are then restricted to the intermediate regime where $G_k M/r \approx 1$.
From Fig.\ \ref{gl3} we see, that these modifications are rather small, however. In particular they do not affect
the horizon structure of the black holes. For the AdS and dS-case the (inner) black hole horizon $\tilde r$ is
slightly shifted to smaller values,
\be
\tilde r_{\rm imp} \lesssim \tilde r_{\rm class} \, , 
\ee
while the cosmological horizon essentially retains its position. The shift of
the inner black hole horizon could be interpreted as shielding of the bare mass parameter due to quantum
gravity corrections. This behavior agrees with earlier findings \cite{Bonanno:2000ep}.
Furthermore, also the
RG-improved Schwarzschild-dS black holes possess a maximal mass
for which the black hole and cosmological horizons coincide. For our illustrative
example this mass agrees with the mass of the classical Nariai mass \eqref{Mcrit}
to very high precision
\be
M_{\rm max,imp}  \simeq \frac{1}{3 G_0 \sqrt{\Lambda_0}} = 10.54 \, , 
\ee
where the same values of $G_0$ and $\Lambda_0$ where used as in previous
graphics.

\begin{figure}[t]
   \centering
\includegraphics[width=10cm]{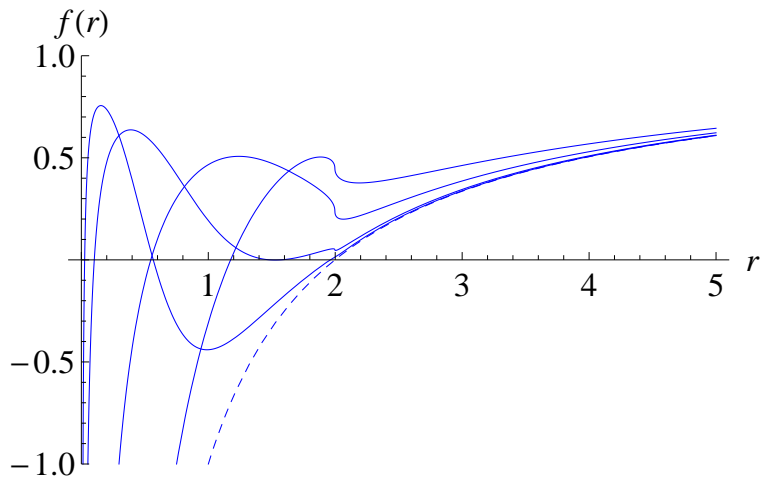}
  \caption{\label{figxidep} 
Functional dependence of the metric function $f(r)$ on the RG-improvement parameter
for a prototypical light Schwarzschild-AdS black hole with $M=1$, $\Lambda_0=-0.001$, $G_0=1$.
From top to bottom, the curves are obtained for  
$\xi=\left\{\xi_{\rm sc}, \, 1.5, \, 0.6, \, 0.3\right\}$. The classical result (dashed line) is included 
for comparison.}
\end{figure}

So far, we have discussed RG-improved black hole solutions for
one particular choice of the cutoff-identification, setting $\xi = \xi_{\rm sc}$.
While this value is distinguished by the self-consistency of the RG-improvement
scheme, it is worthwhile to investigate in as much the results shown in Fig.\ \ref{gl3} depend 
on this choice. Or focus is thereby on ``light'' black holes with mass around the
Planck mass. For a typical Schwarzschild-AdS black hole
the parametric dependence of the improved $f_k(r)$ on $\xi$ is illustrated in 
Fig.~\ref{figxidep}.
The figure illustrates that there is the possibility that the RG-improved solutions
develop an additional pair of horizons if $\xi$ is chosen sufficiently small. These horizons emerge
at intermediate scales $M/r \approx 1$. The asymptotic behavior of the solutions for $M/r \gg 1$ and $M/r \ll 1$
is, however, independent of the choice of $\xi$.  

\begin{figure}[hbt]
   \centering
\includegraphics[width=10cm]{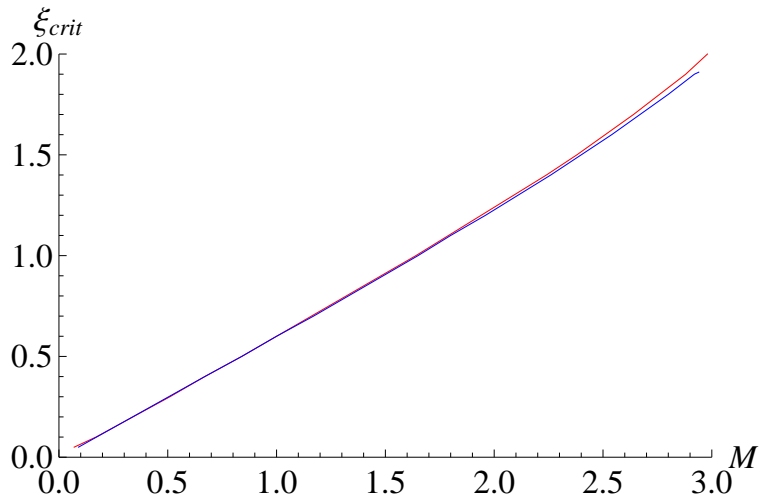}
  \caption{\label{newhorizons} 
Critical value of the RG-improvement parameter $\xi$ for $\Lambda_0 = 0.001$ (red curve) and  $\Lambda_0 = -0.001$ (blue curve) and $G_0 =1$. For $\xi \le \xi_{\rm crit}$ a light RG-improved black hole develops an additional pair of horizons. For masses $M > M_{\rm crit}$ no additional horizons are formed, independently of the actual value of $\xi$.}
\end{figure}

The emergence of a new pair of horizons is investigated more thoroughly in 
Fig.~\ref{newhorizons}. The numerical
study reveals, that this feature only occurs, if the black hole under consideration is sufficiently light, $M \le M_{\rm crit}$.
For the class of Schwarzschild-(A)dS black holes with $\Lambda_0 = \pm 0.001$ the critical mass has been obtained numerically and is given by 
\bea
\mbox{AdS}: &\qquad \Lambda_0 = -0.001& \qquad M_{\rm crit} = 2.94 \\
\mbox{dS}: &\qquad \Lambda_0 = 0.001& \qquad M_{\rm crit} = 2.98 \, .
\eea 
For masses $M \le M_{\rm crit}$ one then finds a critical value of the RG-improvement parameter $ \xi_{\rm crit}(M)$
for which the horizon structure changes. The resulting function $\xi_{\rm crit}(M)$ is shown in Fig.\ \ref{newhorizons}
and depends only weakly on the IR-value of the cosmological constant. 
For black holes with mass $M > M_{\rm crit}$ no additional pair of horizons is formed, independently of
the actual value of the RG-improvement parameter. In particular for the distinguished RG-parameter $\xi_{\rm sc}$
the horizon structure of the RG-improved solution is always the one found in the classical case, independently of
the actual mass of the black hole.

\section{RG-improved black hole solutions: thermodynamics}
\label{sect.7}
After determining the horizon structure of the RG-improved black hole solutions we now
investigate the resulting thermodynamical properties.  We start with discussing the general black hole evaporation process, before investigating the entropy of the microscopic black holes of Sect.\ \ref{sect:7.2} in more detail in Sect.\ \ref{sect:7.1}. Throughout the section we fix $\xi = \xi_{\rm sc}$. 

\subsection{Black hole evaporation}
\label{sect:7.2}
\begin{figure}[t]
   \centering
\includegraphics[width=0.7\textwidth]{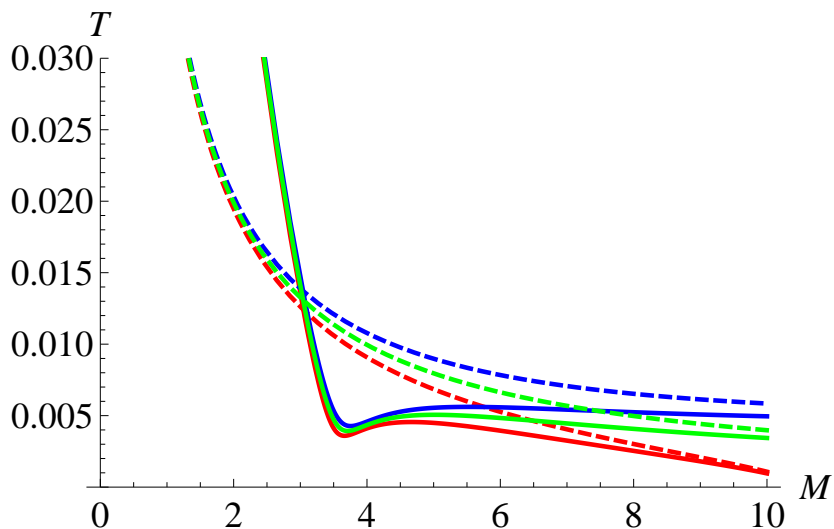}
  \caption{\label{figTempAdS} 
  Temperature of the RG-improved (inner) black hole horizon based on a classical IR value of 
the cosmological constant given by $\Lambda_0 = - 0.001$ (top blue curve), $\Lambda_0 = 0$
(green curve in the middle), and $\Lambda_0 = 0.001$ (red curve bottom) and $G_0 = 1$.
The temperature of the corresponding classical black hole is given by the dashed lines for comparison.}
\end{figure}

The black hole evaporation process is controlled by the inner horizon. The
temperature associated with this horizon is given by \eqref{temp}. For a given
RG-improved solution with fixed $\Lambda_0, G_0$ this expression can easily
be evaluated numerically. The resulting temperature as a function of 
the mass-parameter $M$ is shown in Fig.\ \ref{figTempAdS}. 

The figure illustrates, that for very massive black holes, $M \sqrt{G_0} \gg 1$, the temperature of the classical
and RG-improved solutions actually coincide. The quantum corrections from the RG-improvement set in
once the mass of the black hole becomes comparable to the Planck mass $M \sqrt{G_0} \simeq 1$. In this regime
the temperature of the improved black hole is actually lower than in the classical case, and for a given window situated
around the horizon of the classical black hole, actually decreases with decreasing mass $M$. Below this window
the temperature increases rapidly, so that the sub-Planckian black holes are actually hotter than their classical 
counterparts. For light black holes, the final part of the temperature curve is universal
in the sense that it is completely
controlled by the properties of the NGFP \eqref{NGFP}. In particular it is independent of the IR-value $\Lambda_0$, so
that the final stage of the evaporation process is actually independent whether one considers an asymptotically Schwarzschild or (A)dS black hole.

\begin{figure}[t]
   \centering
\includegraphics[width=0.7\textwidth]{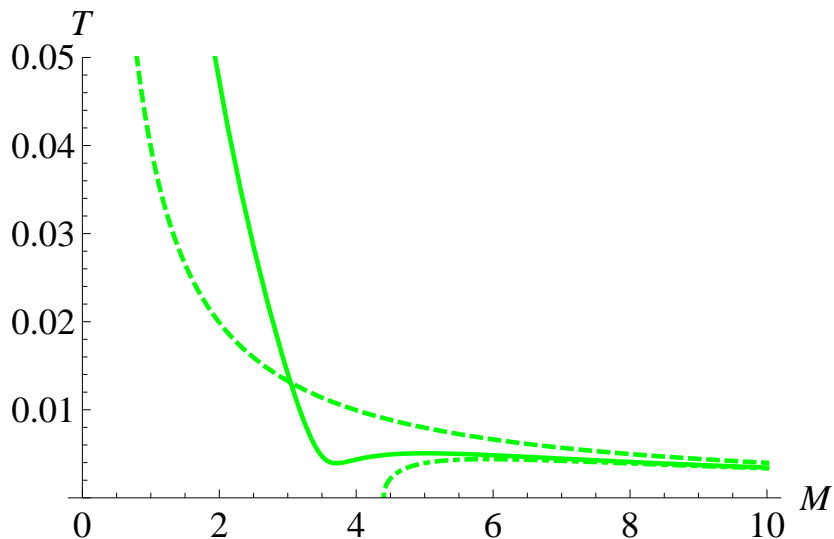} \, 
  \caption{\label{compareAlfioMartin} 
  Comparison between the temperatures of the (RG-improved) Schwarzschild black hole with $\Lambda_0 = 0$ depending on its mass. The classical behavior
is given by the dashed (upper) curve. The RG-improved solution obtained by Bonanno and Reuter \cite{Bonanno:2000ep} (setting $\Lambda_k = 0$ throughout) is given by the dash-dotted (lower) curve
while the horizon temperature obtained from including the running of 
$\Lambda_k$ is given by the solid curve.
}
\end{figure}

At this stage, it is illustrative to compare our findings
with the RG-improved solutions \cite{Bonanno:2000ep},
which approximated $\Lambda_k = 0$ throughout. The temperature of the resulting
RG-improved Schwarzschild black holes are shown in Fig.\ \ref{compareAlfioMartin}.
Comparing the solid curve with the dashed-dotted curve obtained from $\Lambda_k = 0$ it is again
illustrated that the running cosmological constant plays a crucial role for the thermodynamics of
microscopic black holes with mass below the Planck-mass. In particular the inclusion of $\Lambda_k$
prevents that the temperature of the inner horizon drops to zero at a finite mass $M$.
Thus the RG-improved black hole solutions evaporate completely, once the scale-dependence 
of $\Lambda_k$ is taken into account and our analysis indicates that there is no
formation of Planck-mass black hole remnants within Asymptotic Safety.
\subsection{State count for microscopic black holes}
\label{sect:7.1}
We close this section with the computation of the entropy \eqref{entbhh} associated with the inner horizon 
of the quantum-improved Schwarzschild-dS black hole constructed in Sect.\ \ref{SecUVasympt}. Comparing the RG-improved radial function \eqref{fself} and
its classical analogue \eqref{frfct} shows that the entropy of the inner horizon is again
of the form \eqref{horent} with the replacement $G \mapsto G_0 \left(\frac{3}{4} \lambda_* \xi^2 \right)$.
Substituting the inner black hole radius $\tilde r = r_-$, determined by evaluating \eqref{dSroot}
for the RG-improved solution \eqref{fself}, then yields
\be\label{Simp}
\tilde S_* = \frac{4 \pi}{g_* \lambda_*} \, \cos^2\left( \frac{1}{3} \left( \pi + \theta \right)\right) \, . 
\ee  
Here 
\be
\theta \equiv \arctan\left( \sqrt{\frac{4}{27 \tilde m^2} - 1} \right) \, .
\ee
All the RG-improvement scheme dependence can be absorbed into the rescaled mass parameter
\be
\tilde m = M \xi \sqrt{G_0} \, \sqrt{g_*} \, \lambda_* \, . 
\ee
The formulas hold for $0 \le \tilde m^2 \le 4/27$. The upper bound corresponds to the Nariai
black hole with critical mass eq.\ \eqref{Mcrit}. For $\tilde m^2 > 4/27$ the horizons vanish and there is a naked singularity.
This entails that the quantum improved black hole has a $\xi$-independent maximal entropy that is purely determined by
the universal product $g_* \lambda_*$. Setting $\theta = 0$ the maximal entropy of the black hole horizon is
\be\label{smax}
\tilde S_{\rm max} = \, \frac{\pi}{g_* \lambda_*} \, .  
\ee
Notice that, in this limit the black hole and cosmological horizons actually coincide 
so that the total entropy of the black hole is given by twice the black hole horizon entropy.
As a crosscheck, one can use the explicit values of the fixed point coordinate to translate the critical mass
into the physical mass measured in terms of the Planck mass 
\be
M_{\rm max}|_{\xi = \xi_{\rm sc}} = 0.90 M_{\rm Pl} \, .  
\ee
Thus the analysis holds for microscopic black holes with masses around the Planck mass or below.

Most remarkably, the entropy \eqref{smax} can also be obtained by counting the microstates
of the Nariai black holes. This counting can be carried out by exploiting
the `state counting' property of the effective average action \cite{Becker:2012js}
\be
\ln {\mathbb Z}_k \equiv - \Gamma_k[g]|_{g = g_{\rm sol}} \, , 
\ee
where $\Gamma_k[g]$ is to be evaluated at the running self-consistent background. At the level of the
truncation \eqref{action}, $g_{\rm sol}$ is just the solution of the scale-dependent equations of motion resulting
from \eqref{action}. The scale-dependent partition function $\ln {\mathbb Z}_k$ then counts
the field modes integrated out between $k = \infty$ and the IR-scale $k$. 

The Euclidean instanton solutions describing the Nariai black hole
have already been studied in detail \cite{Ginsparg:1982rs,Bousso:1996au,Volkov:2000ih} and correspond to
the Einstein spaces with topology $S^2 \times S^2$. Their 
line-element is given by 
\be\label{euclinst}
ds^2 = \frac{1}{\Lambda_k} \, \bigg( d \theta_1^2 + \sin^2 \theta_1 \, d\varphi_1^2 + d \theta_2^2 + \sin^2 \theta_1 \, d\varphi_2^2 \bigg) \, . 
\ee
Here the scale-dependent cosmological constant reflects that the instanton is the solution 
solves the equations of motion obtained from the scale-dependent $\Gamma_k$. Evaluating the
action \eqref{action} of the instanton, one obtains
\be\label{lnZk}
\ln {\mathbb Z}_k  = - \Gamma_k[g]|_{g = S^2 \times S^2} = \frac{2 \pi}{G_k \, \Lambda_k} \, , 
\ee
where we used that the volume of the unit two-sphere is given by Vol$(S^2) = 4 \pi$. The
entropy \eqref{smax} actually holds on the NGFP. Taking the limit $k \rightarrow \infty$ of (\ref{lnZk}),
we finally obtain
\be\label{statecount}
\ln {\mathbb Z}_* = \frac{2 \pi}{g_* \, \lambda_*} \, . 
\ee
Remarkably, $\ln {\mathbb Z}_* = 2 \tilde S_{\rm max}$ and thus agrees with the total entropy of the two horizons of the Nariai black hole. 
Thus the state counting formula provides a microscopic derivation of the thermodynamic entropy \eqref{smax}. We stress that the result \eqref{statecount}
is actually independent of any RG-improvement procedure and solely 
 relies on the Euclidean instanton action and the existence of the fixed . The matching between the entropy $\tilde S_*$
 and the microstate count suggests that the pertinent
degrees of freedom underlying the thermodynamics of the black hole are the \emph{geometry fluctuations around the background geometry}. 

\section{Summary and Conclusions}
\label{sec.sum}
In this work we have analyzed the structure and thermodynamics of static, spherical symmetric black holes in Quantum Einstein Gravity. In this course we applied the RG-improvement techniques pioneered by Bonanno and Reuter \cite{Bonanno:1998ye,Bonanno:2000ep} and subsequently refined by several groups \cite{Emoto:2005te,Koch:2007yt,Falls:2010he,Casadio:2010fw} to the classical black hole solutions. The crucial novel feature of our analysis is the inclusion of the scale-dependent cosmological constant $\Lambda_k$. As illustrated in Fig.\ \ref{EHflow}, $\Lambda_k$ is always generated dynamically even if the cosmological constant is zero at one particular RG scale. As it was shown in the main text, the inclusion of the running cosmological constant drastically affects on the structure of the quantum-improved black holes \emph{at short distances}. 

At first sight, the modification of short-distance physics due the cosmological constant seems rather counterintuitive since $\Lambda$ is typically connected to gravity phenomena at long distances. In Quantum Einstein Gravity (QEG) the UV-completion of gravity is however provided by a non-Gaussian fixed point of the theories RG flow, implying that the \emph{dimensionless couplings} of the theory become constant at high energies. For the \emph{dimensionful} Newton's constant and cosmological constant this entails
\be\label{fpscaling}
\lim_{k \rightarrow \infty} G _k = g_* k^{-2} \rightarrow 0 \, , \qquad 
\lim_{k \rightarrow \infty} \Lambda_k = \lambda_* k^2 \rightarrow \infty \, ,
\ee
providing a first heuristic argument that at high energies the cosmological constant could be of the same importance as Newtonís constant. Formulated differently, at the non-Gaussian fixed point behaves ``highly quantum'', so that it is a priori unclear that our classical intuition still holds. In fact, our analysis establishes a picture of the quantum-improved black holes which drastically differs from previous QEG studies setting $\Lambda_k = 0$ throughout, even though both rely on the same computational method.

The results of our study of static, spherical symmetric black holes within the Einstein-Hilbert approximation of QEG can be summarized as follows:
\begin{itemize}
\item[a)] {\bf The role of the cosmological constant} \\
Including the effect of a scale-dependent cosmological constant in the RG-improvement process drastically affects the structure of the quantum-improved black holes \emph{at short distances}.
Thus a consistent RG-improvement procedure requires working in the class of Schwarzschild-(A)dS solutions of Einsteinís equations.
\item[b)] {\bf Universal short distance behavior} \\
The short-distance structure of all quantum-improved black holes is governed by the non-Gaussian RG fixed point. This entails that the structure of light black holes is universal. In particular it is independently of the IR-value of Newtonís constant and the cosmological constant and therefore identical for classical Schwarzschild, Schwarzschild-dS and Schwarzschild-AdS black holes. 
\item[c)] {\bf Curvature singularities} \\
In the presence of the cosmological constant, the curvature singularity at $r = 0$ is not resolved.
\end{itemize}
The results a) to c) hold independent of the actual RG-improvement scheme. Moreover, we made the following,
RG-improvement scheme-dependent observations:
\begin{itemize}
\item[d)] {\bf Distinguished RG-improvement scheme} \\
In the main text we focused on a particular RG-improvement scheme (initially proposed in \cite{Bonanno:2000ep}) which relates the RG-scale $k$ to the radial geodesic distance of a point $P$ to the origin. This scheme is singled out by the fact that it is self-consistent in the UV: Both the classical and RG-improved solutions give rise to the same cutoff-identification. 
Moreover, results are consistent with applying the RG-improvement at the level of the equations of motion and flowing action $\Gamma_k$
and to the alternative choice of geodesic distance examined in \ref{App.0}.
\item[e)] {\bf Classical-quantum black hole duality}\\
Working within this distinguished RG-improvement scheme, the quantum-improved microscopic black holes
are described by the \emph{classical Schwarzschild-dS solution}. In the improvement process
the cosmological constant and Newton's constant actually swap places, so that it is $\lambda_*$ that
actually controls the short distance behavior of the solution while $g_*$ takes over the role 
of the classical cosmological constant term.
\item[f)] {\bf Black hole evaporation and remnant formation} \\
The inclusion of the cosmological constant leads  
to a complete evaporation
of the asymptotically safe black holes. In particular
there is no mechanism any more that enforces the formation of
Planck-mass remnants which implies that naked singularities formed by the radiating quantum-improved black holes. The final stage of the evaporation process
 is universal and governed by the properties of the non-Gaussian fixed point
which determines the temperature of the inner horizon of the microscopic black holes. 

\item[g)] {\bf Microstate count} \\
For the Nariai black hole, appearing in a particular limit of the quantum-improved black hole solutions in the deep UV, the thermodynamic entropy of the black hole configuration agrees with the microscopic state count based on the effective average action. This matching suggests that the degrees of freedom responsible for the black hole entropy are quantum fluctuations around the mean-field black hole configuration and constitutes the first microscopic derivation of a thermodynamical entropy within Asymptotic Safety.
\end{itemize}
Notably, all of these conclusions remain valid when performing the proper-time cutoff identification advocated in \ref{App.0}, adding further evidence for their robustness.

In particular the result e) opens up new perspectives on black holes in Asymptotic Safety. 
On one hand the emerging picture is reminiscent of the classicalization scenario advocated in \cite{Dvali:2010bf}, which proposes that the effective degrees of freedom of gravity at high energies are given by classical black hole solutions. The observation that the asymptotically safe black holes are self-similar in the sense that the quantum-improved black holes found by evaluating the RG-improvement at the non-Gaussian fixed point are again described by the classical Schwarzschild-dS solution may hint at a hitherto unnoticed connection between these ideas.

A second profound consequence of the observation that the quantum-improved black holes found at the non-Gaussian fixed point are actually Schwarzschild-dS is that their entropy can be expressed in terms of the Cardy-Verlinde formula \cite{Cardy:1986ie}, which is conjectured to describe the entropy of a CFT in general dimension \cite{Verlinde:2000wg}. This, in particular, dismisses the arguments \cite{Shomer:2007vq} that entropy count of Asymptotic Safety is not compatible with the one expected from a CFT. Moreover, it opens up the exciting possibility that there is a description of the quantum-black hole geometry on the basis of the dS/CFT correspondence \cite{Strominger:2001pn}. 

Finally, we expect that the
analysis of more complex black hole solutions
including electromagnetic charges and angular momentum will
also be affected by our findings. In particular, the 
 evaporation of asymptotically safe black holes in theories with large extra
dimensions \cite{Hewett:2007st,Litim:2007iu,Koch:2007yt,Burschil:2009va} may be altered by the inclusion of the cosmological
constant. We leave the investigation of these point to future works.

\section*{Acknowledgements}
We thank M.\ Reuter for helpful discussions and an anonymous referee 
for suggesting the proper time improvement studied in \ref{App.0}.
F.S.\ thanks
the Pontificia Universidad Cat\'olica de Chile (Santiago) for 
hospitality during the initial stage of the project. 
The work of B.K.\ was supported proj.\ Fondecyt 1120360 
and anillo Atlas Andino 10201 while the research of F.S.\ is
supported by the Deutsche Forschungsgemeinschaft (DFG)
within the Emmy-Noether program (Grant SA/1975 1-1).

\begin{appendix}
\section{Proper time RG improvement}
\label{App.0}
In this section we will study an 
alternative choice for the scale-setting
procedure (\ref{kvonP}). In this case, the scale  
$k$ is identified with the inverse of the proper time 
measured by a freely falling observer starting at rest at $r_0$
and reaching the classical black hole singularity after the time $\tau$.
We will call this novel RG-improvement scheme proper-time improvement (pti).
The main advantage of this procedure is that it leads to a cutoff identification
that is smooth when passing through a horizon, thereby eliminating the points with infinite slope appearing
in Figure \ref{gl2}. As the main result of this appendix, we establish that all conclusions
listed in the summary are confirmed by this alternative RG-improvement scheme.

We start by deriving the proper time interval $\tau(r_0)$ passing on the clock of a freely falling observer
starting at rest at $r_0$ before reaching the black hole singularity at $r=0$. For the Schwarzschild solution
this computation can, e.g., be found in \cite{HartleSolutions}
and it is straight forward to generalize it to the case with 
a cosmological constant. Starting from the geodesic equation
derived from the classical line element \eqref{frfct}, one first establishes
from the time component that
\be
E=(1-\frac{2GM}{r}-\frac{\Lambda r^2}{3}) u^0 \, , 
\ee
with $u^\nu=dx^\nu/d\tau$ denoting the observers four-velocity, is a conserved 
quantity. Evaluated for the observer at rest at $r_0$ the energy can be computed 
explicitly
\be
E^2 = (1-\frac{2GM}{r_0}-\frac{\Lambda r^2_0}{3}) \, .
\ee
Using the normalization condition of the four-velocity, $u^\nu u_\nu=-1$,
this gives rise to an equation for the radial velocity 
\be
\frac{d r}{d\tau}=\sqrt{\left(1-\frac{2GM}{r_0}-\frac{\Lambda}{3}r_0^2\right)-\left(1-\frac{2GM}{r}-\frac{\Lambda}{3}r^2\right)}\, .
\ee
The proper time $\tau(r_0)$ depending on the starting point of the observer 
is obtained by integrating this relation
\be\label{geoddist}
\tau(r_0) = \int_0^{r_0} dr \,(  2 G
M/r +  1/3 \Lambda r^2 - 2 G M/r_0 - 1/3 \Lambda r_0^2 )^{-1/2}\, .
\ee
The proper-time improvement then uses this expression to related the cutoff scale $k(P(r))$, eq.\ \eqref{kvonP},
to the inverse of this expression
\be\label{pti}
k(P(r)) = \frac{\xi}{\tau(r)} \, . 
\ee

At this stage, the following remark is in order. For the AdS and Schwarzschild
black hole ($\Lambda \le 0$) this scale-setting procedure provides a good cutoff identification
in the sense that $r \rightarrow 0$ corresponds to $k \rightarrow \infty$ and vice versa.
 For dS black holes, however, the integral
\eqref{geoddist} diverges at a finite radius $r_{\rm tp}=(3 G M/\Lambda)^{1/3}$ which is located between
the black hole and cosmological horizon of the geometry.
Observers starting at $r_0 > r_{\rm tp}$ will not end up in the central black hole singularity. Instead
they are falling outward towards the cosmological horizon. Thus, in this case we still have that at short distances
$r \rightarrow 0$ implies $k(r) \rightarrow \infty$ while in the IR $k(r) \rightarrow 0$ as $r \rightarrow r_{\rm tp}$. The whole RG-trajectory 
is therefore evaluated in a finite coordinate interval $r \in [0, r_{\rm tp}]$. This suggests that 
the proper-time improvement does not lead to a valid cutoff identification in the dS case and we restrict
our analysis of global properties to the AdS and Schwarzschild black holes.

When investigating the phenomenological consequences arising from the proper-time improvement scheme,
we start with the analysis for small radii. In this case (\ref{geoddist}) can be applied to both
 the dS and the AdS case. Expanding the integrand of equation (\ref{geoddist})
the short distance asymptotics of $\tau(r)$ is given by
\be
\tau(r) = \frac{\pi}{2} \, \frac{1}{\sqrt{2 G M}} \, r^{3/2} \,  ( 1 + {\mathcal{O}}(r)) \, .
\ee
\begin{figure}[t]
  \centering
\subfloat[\hspace{7cm}.
 Proper-time cutoff identification $k_{pti}(r)/\xi$ obtained for the AdS black holes with $M=\{1,\, 2.5, \,5, \, 10\}$ (bottom to top).
 ]{\label{figa1}
\includegraphics[width=0.48\textwidth]{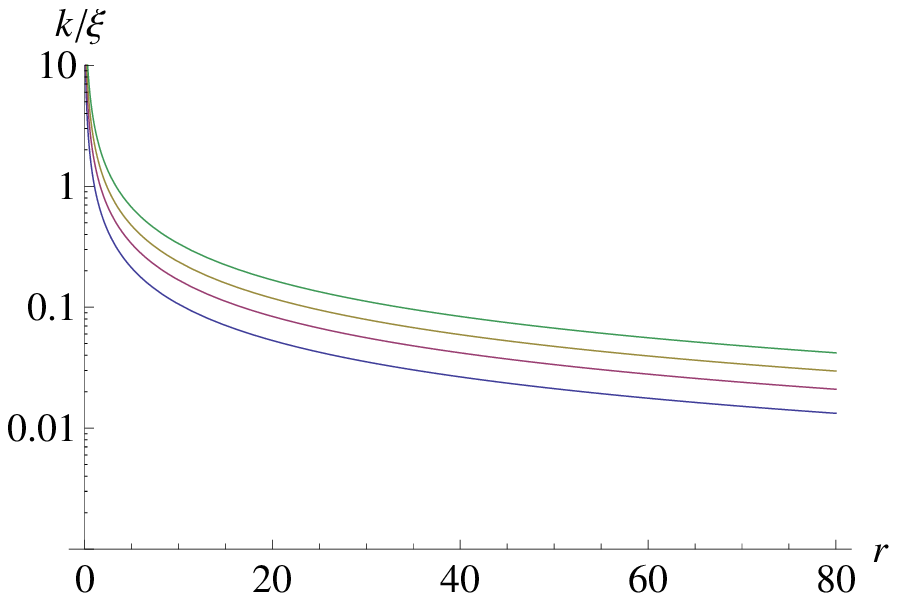}}
\hspace{0.2cm}
\subfloat[\hspace{7.0cm}.
 The RG-improved metric function
 $f(r)$ for the AdS black holes with mass $M=\{1,\, 2.5, \,5, \, 10\}$ (top to bottom) and $\xi = \xi_{\rm sc}^{\rm pti}$. The dashed lines
 illustrate the corresponding classical case.
 ]{\label{figa2}
\includegraphics[width=0.48\textwidth]{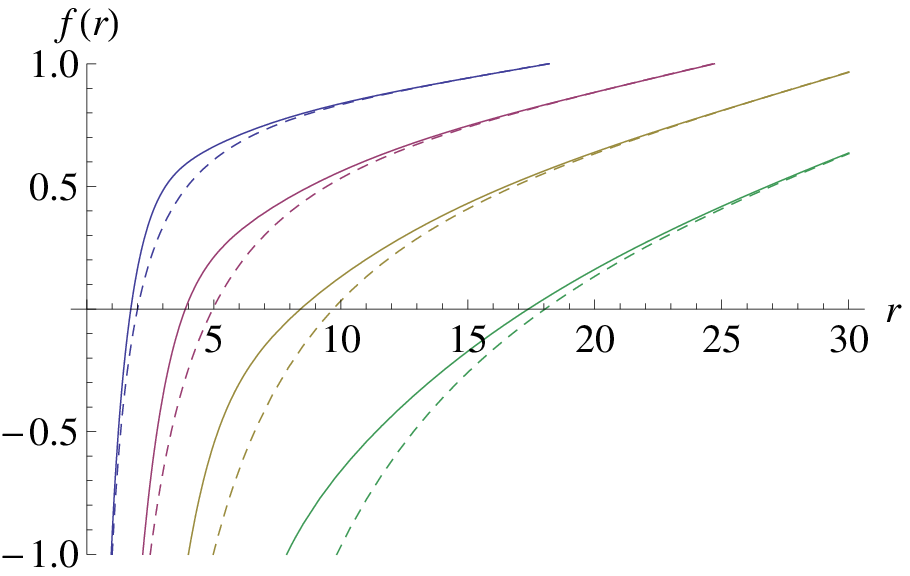}}
\caption{Functional dependence of the scale
  $k_{pti}(r)/\xi$ (left) and the resulting RG-improved metric factor $f(r)$ (right) for Schwarzschild-AdS black holes with $\Lambda_0=-0.001$ and $G_0 = 1$.
 \label{figvonrgeo}}
\label{fa3}
\end{figure}
Comparing this result with the short distance expansion (\ref{radasym}) shows that both
RG-improvement schemes give rise to the \emph{same short distance behavior}. They only differ
in a numerical factor which can be reabsorbed in a redefinition of $\xi$. Thus the UV approximation
obtained from the proper-time improvement is identical to the one discussed in section \ref{SecUVasympt},
lending further support to the robustness of the results obtained in the main text. In particular
there is again a self-consistent solution where
\be
\xi_{\rm sc}^{\rm pti} = \frac{3 \pi}{4} \xi_{sc} \, . 
\ee

For the AdS and Schwarzschild black holes, it also makes sense to discuss the 
global global properties of the RG-improved black hole solutions along the lines
of section \ref{SecUVIR}. The cutoff identification $k(r)$ arising from
the proper-time improvement is shown in the left diagram of figure \ref{fa3}.
Comparing the result to the cutoff identification using the proper distance shown in figure \ref{gl2} establishes that the schemes give rise to the same asymptotic behavior for small and large values $r$. Their only difference arises when $r$ is close to the black hole horizon: here the proper-time improvement gives rise to an infinite slope which is smoothed out in the proper-time scheme. As a direct consequence,
we find that the RG-improved metric function $f(r)$ for $\xi = \xi_{\rm sc}^{\rm pti}$, shown in the right diagram of figure \ref{figvonrgeo} remains monotonic. Decreasing $\xi$, however, we find that also the proper-time improvement scheme can give rise to additional horizon pairs. In this respect, we recover the global structure of the RG-improved black hole solutions described in section \ref{SecUVIR}.

Following the strategy of section \ref{sect.7}, it is also straight forward to obtain the horizon temperature of the RG-improved AdS black hole arising from the proper-time improvement.
\begin{figure}[t]
\centering
\includegraphics[width=10cm]{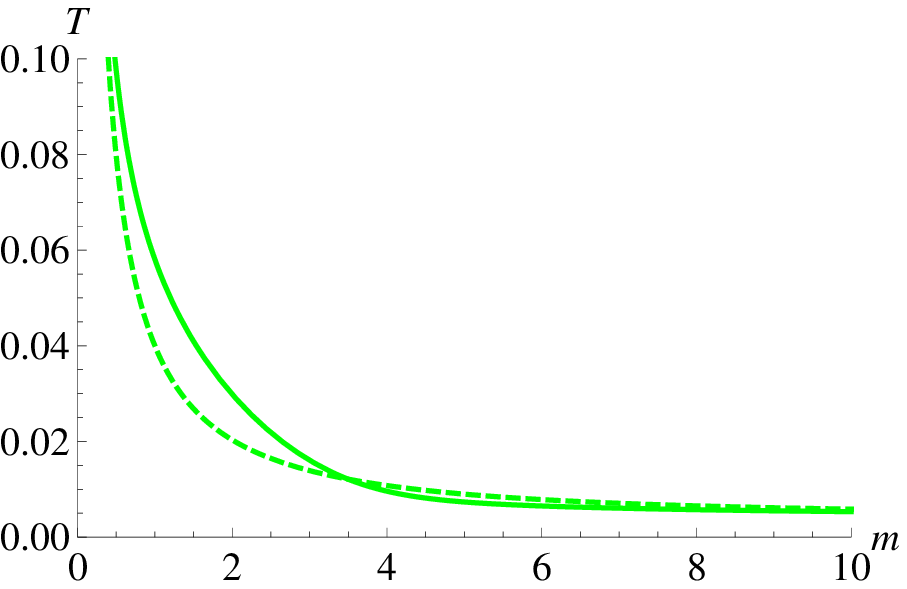}
\caption{\label{temppti} 
Temperature of the RG-improved AdS black hole with $\Lambda_0 = -0.001$ obtained from the proper-time improvement scheme with $\xi = \xi^{\rm pti}_{\rm sc}$. The temperature of the corresponding classical black hole
is given by the dashed line for comparison.}
\end{figure}
The result, shown in Figure \ref{temppti}, closely resembles the temperature curve obtained from the RG-improvement based on the proper distance shown in Figure \ref{figTempAdS}. 
In particular the asymptotics for very small and large masses $m$ are identical. One noticeable difference is that the proper-time improvement leads to a heat-capacity that is strictly negative definite.
Since the proper-time improvement is precisely tailored for improving the RG-improvement close to the horizon, we expect, that this actually reflects a genuine property of the RG-improved black hole that is not captured by the proper-distance improvement.  

To summarize the results of this appendix, we have found that \emph{all conclusions} summarized in
section \ref{sec.sum} are also valid for the proper-time improvement scheme. This provides additional
evidence for the robustness of the results derived in this work.

\section{Improving equations of motion and action}
\label{App.I}
In this appendix we establish that the quantum-improved
solutions constructed in Sect.\ \ref{SecUVasympt} can also be
obtained from other RG-improvement schemes. Their compatibility with
 ``improving the equations of motion'' \cite{Bonanno:2001hi}
and ``improving the action'' \cite{Reuter:2003ca} will be
demonstrated in \ref{App.A1} and \ref{App.A2}, respectively.

\subsection{Improving the equations of motion}
\label{App.A1}
An alternative route for extracting phenomenological
consequences from the running effective average action
is the RG-improvement of the equations of motion. In this
case the cutoff-identification $k \mapsto k(r)$ is applied
at the level of the scale-dependent equations of 
motion derived from $\Gamma_k$. As a consequence,
the coupling constants become position-dependent
$G_k \mapsto G(x)$ with the dynamics provided by
the RG-equations.

For the Einstein-Hilbert ansatz \eqref{action}
the RG-improved equations of motion
read \cite{Reuter:2004nx}
\begin{eqnarray}
 G_{\mu\nu}=-g_{\mu\nu}\Lambda_{k}+8\pi G_{k}T_{\mu\nu}-
\Delta t_{\mu \nu}\, , 
\label{eom}
\end{eqnarray}
with $G_{\mu\nu} \equiv R_{\mu\nu} - \frac{1}{2} g_{\mu\nu} R$ being the Einstein-tensor. The position-dependent Newton's constant
leads to an
 additional contribution to the stress energy tensor
\cite{Reuter:2004nx,Carroll:2004st} 
\be\label{deltatmunu}
\Delta t_{\mu \nu}=G_{
k}\left(g_{\mu\nu} \, D^2-D_\mu D_\nu\right)\frac{1}{G_{k}} \, .
\ee 
The consistency of eq.\ \eqref{eom} then allows
to determine the cutoff-identification $k = k(r)$
dynamically. Assuming the conservation of the
stress-energy tensor
\be\label{Tmunu}
T_{\mu\nu}^{\quad;\nu}=0 
\ee
and applying $g_{\mu\nu}^{\quad;\nu}=0$ together with the Bianchi identity
\begin{eqnarray}
 G_{\mu\nu}^{\quad;\nu}=0 \, ,
\label{Bianchi}
\end{eqnarray}
eq.\ \eqref{eom} implies that \cite{Koch:2010nn}
\be\label{cond}
R\, D_\mu \left(\frac{1}{G_k}\right)-
2 \, D_\mu\left(\frac{\Lambda_k}{G_k}\right)=0 \, .
\ee
For a given form of the running couplings this equation
can be used to determine the cutoff-identification $k(R)$.
In general this relation will include arbitrary powers
of the curvature which make the equations of motion (\ref{eom}) hard to handle.
However, this situation simplifies in the deep UV
where the scale-dependence of the couplings
simplifies to \eqref{UVG}. Substituting the scaling
relation at the NGFP, \eqref{cond} becomes
\be\label{condfp}
\frac{\lambda_*}{g_*} \left( R - 4 \lambda_* k^2 \right) \left( D_\mu k^2 \right) = 0 \, . 
\ee
Asking for a non-trivial solution where $D_\mu k^2 \not = 0$ 
provides the desired relation between $k^2$ and
the scalar curvature
\be\label{kvonR}
k^2=\frac{R}{4 \lambda_*} \, .
\ee
This confirms the linear dependency between
$R$ and $k^2$ as it is often used in the FRGE literature and 
also fixes the constant that comes with this proportionality.

We now focus on vacuum solutions of \eqref{eom}, setting $T_{\mu\nu} = 0$.
Substituting the fixed point scaling \eqref{UVG} and subsequently
performing the cutoff-identification \eqref{kvonR}, we obtain
the improved equations of motion
\emph{valid in the scaling regime of the NGFP}
\be\label{eomUV}
R_{\mu \nu}-\frac{1}{4}g_{\mu \nu} R+ \frac{1}{R}\left(g_{\mu\nu}D^2-D_\mu D_\nu\right)R=0 \, .
\ee
At this stage, it is straightforward to check that \eqref{fself} also solves these
 RG-improved equations of motion. Thus we verified that
our RG-improved fixed point solution is also compatible with the ``improved equations of motion''-scheme.

\subsection{Improving the action}
\label{App.A2}
The RG-improvement procedure can also
be applied at the level of the scale-dependent
Lagrangian density appearing in the effective
average action. This procedure is called
``improving the action''. We now demonstrate
that the solution \eqref{fself} can also be
obtained from this RG-improvement method.

At the NGFP, the (lorentzian) effective
average action \eqref{action} takes
the form
\be\label{fpaction}
\Gamma_*[g] = \frac{1}{16 \pi} \int d^4x \sqrt{-g} \, k^2 \, g_*^{-1} \, \left( R - 2 \lambda_* k^2 \right) 
\ee
Following the discussion \cite{Bonanno:2012jy}, a natural choice
for the cutoff-identification is to postulate a linear relation between
the renormalization scale and the curvature,
\be
k^2= \, \xi \, R \, , 
\ee
which generalizes \eqref{kvonR} by including an arbitrary
proportionality constant. Substituting this
relation into \eqref{fpaction}, the RG-improved
action becomes quadratic in the curvature scalar
\be
\Gamma_{*}=\frac{\xi}{16\pi g_*} \, \left(1- 2 \lambda_* \xi\right) \, \int d^4 x \, \sqrt{-g} \, R^2 \, .
\ee
For any value of $\xi \neq 1/(2 \lambda_*)$, the equations of motion
resulting from this action are again given by \eqref{eomUV}. This shows the internal consistency of ``improving the equations of motion''
and ``improving the action''. In particular the fixed point solution \eqref{fself}
can be obtained from both procedures.
We stress that this consistency would not hold
if the cutoff-identification \eqref{dproper} is replaced by setting $k = \xi/r$.

We close this appendix with the following comments.
Recently a special solution of \eqref{eom} has been found
in terms of the metric and the radius dependent fields $G(r)$ and $\Lambda(r)$ \cite{Contreras:2013hua}.
Whether this solution really reflects effects due to running couplings is however not yet clear.
In this light, tracing \eqref{eomUV} gives
 \be\label{treomUV}
 3\frac{\Box R}{R} =0 \, ,
 \ee
 which might be useful for finding more general solutions of (\ref{eomUV}).
Finally, one might also consider the improved equations
 of motion (\ref{eom}) without the contribution (\ref{deltatmunu}). 
This is however not in the spirit of this work since is would imply
 a non-zero and non-conserved stress energy tensor for matter.

\bigskip \bigskip \bigskip
\end{appendix}



\begin{thebibliography}{15}

\bibitem{Carroll:2004st}
  S.~M.~Carroll,
  \emph{Spacetime and geometry: An introduction to general relativity},
  Addison-Wesley, San Francisco, USA, (2004).
  
\bibitem{Wald:1995yp} 
  R.~M.~Wald,
  \emph{Quantum field theory in curved space-time and black hole thermodynamics},
  Chicago Univ. Press, Chicago, USA, (1994).



\bibitem{mr}
M.~Reuter,
Phys.\ Rev.\ D {\bf 57}, 971 (1998), hep-th/9605030.

\bibitem{Weinberg:1979}
S.~Weinberg
in \textit{General Relativity, an Einstein Centenary Survey},
S.W.~Hawking and W.~Israel (Eds.),
Cambridge University Press, 1979; 
S.~Weinberg, hep-th/9702027.




\bibitem{Niedermaier:2006wt} 
  M.~Niedermaier and M.~Reuter,
  Living Rev.\ Rel.\  {\bf 9}, 5 (2006).

\bibitem{Reuter:2007rv} 
M.~Reuter and F.~Saueressig, in {\it Geometric and Topological Methods for Quantum Field Theory}, H.~Ocampo, S.~Paycha and
 A.~Vargas (Eds.), Cambridge Univ.\ Press, Cambridge (2010); arXiv:0708.1317.

\bibitem{Percacci:2007sz}
  R.~Percacci,
   R.~Percacci, in \textit{Approaches to Quantum Gravity: Towards a New Understanding of Space, Time and Matter}, D. Oriti (Ed.), Cambridge Univ.\ Press, Cambridge (2009);
  arXiv:0709.3851.

\bibitem{Litim:2008tt}
  D.~F.~Litim, 
  PoS(QG-Ph) {\bf 024} (2008);
  arXiv:0810.3675.

\bibitem{Reuter:2012id} 
  M.~Reuter and F.~Saueressig,
  New J.\ Phys.\  {\bf 14}, 055022 (2012);
  arXiv:1202.2274.




\bibitem{Dou:1997fg}
  D.~Dou and R.~Percacci,
  Class.\ Quant.\ Grav.\  {\bf 15}, 3449 (1998);
  hep-th/9707239.

\bibitem{Souma:1999at}
  W.~Souma,
  Prog.\ Theor.\ Phys.\  {\bf 102}, 181 (1999);
  hep-th/9907027.

\bibitem{Lauscher:2001ya} 
  O.~Lauscher and M.~Reuter,
  Phys.\ Rev.\ D {\bf 65}, 025013 (2002);
  hep-th/0108040.

\bibitem{Reuter:2001ag}
  M.~Reuter and F.~Saueressig,
  Phys.\ Rev.\  D {\bf 65}, 065016 (2002);
  hep-th/0110054.

\bibitem{Litim:2003vp}
  D.~F.~Litim,
  Phys.\ Rev.\ Lett.\  {\bf 92}, 201301 (2004);
  hep-th/0312114.


\bibitem{Lauscher:2001rz} 
  O.~Lauscher and M.~Reuter,
  Class.\ Quant.\ Grav.\  {\bf 19}, 483 (2002);
  hep-th/0110021.

\bibitem{Lauscher:2002sq} 
  O.~Lauscher and M.~Reuter,
  Phys.\ Rev.\ D {\bf 66}, 025026 (2002);
  hep-th/0205062.



\bibitem{Rechenberger:2012pm} 
  S.~Rechenberger and F.~Saueressig,
  Phys.\ Rev.\ D {\bf 86}, 024018 (2012);
  arXiv:1206.0657.

\bibitem{Codello:2007bd}
  A.~Codello, R.~Percacci and C.~Rahmede,
  Int.\ J.\ Mod.\ Phys.\  A {\bf 23}, 143 (2008);
  arXiv:0705.1769.

\bibitem{Machado:2007ea} 
  P.~F.~Machado and F.~Saueressig,
  Phys.\ Rev.\ D {\bf 77}, 124045 (2008);
  arXiv:0712.0445.

\bibitem{Bonanno:2010bt} 
  A.~Bonanno, A.~Contillo and R.~Percacci,
  Class.\ Quant.\ Grav.\  {\bf 28}, 145026 (2011);
  arXiv:1006.0192.

\bibitem{Falls:2013bv} 
  K.~Falls, D.~F.~Litim, K.~Nikolakopoulos and C.~Rahmede,
  arXiv:1301.4191.




\bibitem{Christiansen:2012rx} 
  N.~Christiansen, D.~F.~Litim, J.~M.~Pawlowski and A.~Rodigast,
  arXiv:1209.4038.

\bibitem{Benedetti:2009rx} 
  D.~Benedetti, P.~F.~Machado and F.~Saueressig,
  Mod.\ Phys.\ Lett.\ A {\bf 24}, 2233 (2009);
  arXiv:0901.2984.

\bibitem{Benedetti:2009gn} 
  D.~Benedetti, P.~F.~Machado and F.~Saueressig,
  Nucl.\ Phys.\ B {\bf 824}, 168 (2010);
  arXiv:0902.4630.

\bibitem{Niedermaier:2009zz} 
  M.~R.~Niedermaier,
  Phys.\ Rev.\ Lett.\  {\bf 103}, 101303 (2009).


\bibitem{Eichhorn:2009ah} 
  A.~Eichhorn, H.~Gies and M.~M.~Scherer,
  Phys.\ Rev.\ D {\bf 80}, 104003 (2009);
  arXiv:0907.1828;
  K.~Groh and F.~Saueressig,
  J.\ Phys.\ A {\bf 43}, 365403 (2010);
  arXiv:1001.5032;
  A.~Eichhorn and H.~Gies,
  Phys.\ Rev.\ D {\bf 81}, 104010 (2010);
  arXiv:1001.5033.
  A.~Eichhorn,
  arXiv:1301.0632.


\bibitem{Becker:2012js} 
  D.~Becker and M.~Reuter,
  JHEP {\bf 1207}, 172 (2012), arXiv:1205.3583. D.~Becker and M.~Reuter, arXiv:1212.4274.



 
\bibitem{Bonanno:1998ye}
  A.~Bonanno and M.~Reuter,
  Phys.\ Rev.\  D {\bf 60}, 084011 (1999);
  gr-qc/9811026.

\bibitem{Bonanno:2000ep}
  A.~Bonanno and M.~Reuter,
  Phys.\ Rev.\  D {\bf 62}, 043008 (2000);
  hep-th/0002196.

\bibitem{Emoto:2005te} 
  H.~Emoto,
  hep-th/0511075.

\bibitem{Bonanno:2006eu} 
  A.~Bonanno and M.~Reuter,
  Phys.\ Rev.\ D {\bf 73}, 083005 (2006);
  hep-th/0602159.

\bibitem{Ward:2006vw} 
  B.~F.~L.~Ward,
  Acta Phys.\ Polon.\ B {\bf 37}, 1967 (2006);
  hep-ph/0605054.

\bibitem{Falls:2010he} 
  K.~Falls, D.~F.~Litim and A.~Raghuraman,
  Int.\ J.\ Mod.\ Phys.\ A {\bf 27}, 1250019 (2012);
  arXiv:1002.0260.


\bibitem{Basu:2010nf} 
  S.~Basu and D.~Mattingly,
  Phys.\ Rev.\ D {\bf 82}, 124017 (2010);
  arXiv:1006.0718.

\bibitem{Contreras:2013hua} 
  C.~Contreras, B.~Koch and P.~Rioseco,
  arXiv:1303.3892.



\bibitem{Reuter:2006rg} 
  M.~Reuter and E.~Tuiran,
  hep-th/0612037.

\bibitem{Reuter:2010xb}
  M.~Reuter and E.~Tuiran,
  Phys.\ Rev.\ D {\bf 83}, 044041 (2011);
  arXiv:1009.3528.


\bibitem{Cai:2010zh} 
  Y.-F.~Cai and D.~A.~Easson,
  JCAP {\bf 1009}, 002 (2010);
  arXiv:1007.1317.
  

\bibitem{Falls:2012nd} 
  K.~Falls and D.~F.~Litim,
  arXiv:1212.1821.



\bibitem{Hewett:2007st} 
  J.~Hewett and T.~Rizzo,
  JHEP {\bf 0712}, 009 (2007);
  arXiv:0707.3182.

\bibitem{Litim:2007iu} 
  D.~F.~Litim and T.~Plehn,
  Phys.\ Rev.\ Lett.\  {\bf 100}, 131301 (2008);
  arXiv:0707.3983.

\bibitem{Koch:2007yt}
  B.~Koch,
  Phys.\ Lett.\  B {\bf 663}, 334 (2008);
  arXiv:0707.4644.


\bibitem{Burschil:2009va}
  T.~Burschil and B.~Koch,
  Zh.\ Eksp.\ Teor.\ Fiz.\  {\bf 92}, 219 (2010);
  arXiv:0912.4517.




\bibitem{Casadio:2010fw}
  R.~Casadio, S.~D.~H.~Hsu and B.~Mirza,
  Phys.\ Lett.\ B {\bf 695}, 317 (2011);
  arXiv:1008.2768.



\bibitem{Cai:2001sn}
  R.-G.~Cai,
  Phys.\ Lett.\ B {\bf 525} (2002) 331;
  hep-th/0111093.

\bibitem{Cai:2001tv}
  R.-G.~Cai,
  Nucl.\ Phys.\ B {\bf 628} (2002) 375;
  hep-th/0112253.
  
\bibitem{Verlinde:2000wg} 
  E.~P.~Verlinde,
  hep-th/0008140.

\bibitem{Balasubramanian:2001nb} 
  V.~Balasubramanian, J.~de Boer and D.~Minic,
  Phys.\ Rev.\ D {\bf 65}, 123508 (2002);
  hep-th/0110108.
  
\bibitem{Ghezelbash:2001vs} 
  A.~M.~Ghezelbash and R.~B.~Mann,
  JHEP {\bf 0201}, 005 (2002);
  hep-th/0111217.
  
\bibitem{Abbott:1981ff} 
  L.~F.~Abbott and S.~Deser,
  Nucl.\ Phys.\ B {\bf 195}, 76 (1982).

\bibitem{Koch:2010nn} 
  B.~Koch and I.~Ramirez,
  Class.\ Quant.\ Grav.\  {\bf 28}, 055008 (2011);
  arXiv:1010.2799.

\bibitem{Reuter:2012xf} 
M.\ Reuter and F.\ Saueressig,  Lect.\ Notes Phys.\ {\bf 863}, 185 (2013);
 arXiv:1205.5431. 

 
\bibitem{Ginsparg:1982rs} 
  P.~H.~Ginsparg and M.~J.~Perry,
  Nucl.\ Phys.\ B {\bf 222}, 245 (1983).

\bibitem{Bousso:1996au} 
  R.~Bousso and S.~W.~Hawking,
  Phys.\ Rev.\ D {\bf 54}, 6312 (1996);
  gr-qc/9606052.

\bibitem{Volkov:2000ih} 
  M.~S.~Volkov and A.~Wipf,
  Nucl.\ Phys.\ B {\bf 582}, 313 (2000);
  hep-th/0003081.

\bibitem{Dvali:2010bf} 
  G.~Dvali and C.~Gomez,
  arXiv:1005.3497.

\bibitem{Cardy:1986ie} 
  J.~L.~Cardy,
  Nucl.\ Phys.\ B {\bf 270}, 186 (1986).
  
\bibitem{Shomer:2007vq} 
  A.~Shomer,
  arXiv:0709.3555.

  
\bibitem{Strominger:2001pn} 
  A.~Strominger,
  JHEP {\bf 0110}, 034 (2001);
  hep-th/0106113.
  
  \bibitem{Bonanno:2001hi}
  A.~Bonanno and M.~Reuter,
  Phys.\ Lett.\  B {\bf 527}, 9 (2002);
  astro-ph/0106468.
  A.~Bonanno and M.~Reuter,
  Phys.\ Rev.\  D {\bf 65}, 043508 (2002);
  hep-th/0106133.

\bibitem{Reuter:2003ca}
  M.~Reuter and H.~Weyer,
  Phys.\ Rev.\  D {\bf 69}, 104022 (2004);
  hep-th/0311196.
  
\bibitem{Reuter:2004nx}
  M.~Reuter and H.~Weyer,
  JCAP {\bf 0412}, 001 (2004);
  hep-th/0410119.
  
\bibitem{Bonanno:2012jy} 
  A.~Bonanno,
  Phys.\ Rev.\ D {\bf 85}, 081503 (2012);
  arXiv:1203.1962.
 
 \bibitem{HartleSolutions}
 J.~B.~Hartle, Gravity, An Introduction to General Relativity, Solutions to Problems, Problem 12.5 (2003);
 ISBN 0-8053-8663-7 .
  
\end{thebibliography}
\end{document}